\documentclass[aps,prl,twocolumn,showpacs,superscriptaddress,preprintnumbers,amsmath,amssymb,longbibliography]{revtex4-1}

\usepackage{dcolumn}
\usepackage{bm}
\usepackage[normalem]{ulem}    
\usepackage{xcolor}
\usepackage{gensymb}   
\usepackage{tikz}
\usepackage{physics}
\usepackage[pdftex,colorlinks=true,pdfstartview=FitV,linkcolor=blue,citecolor=blue,urlcolor=blue]{hyperref}
\usepackage{tabu}

\newcommand{\kk}{\mathbf{k}}

\begin{document}

\title{Emergence of flat bands in the quasicrystal limit of boron nitride twisted bilayers}
\author{Lorenzo Sponza}
\affiliation{Universit\'{e} Paris-Saclay, ONERA, CNRS, Laboratoire d'\'{e}tude des microstructures (LEM), 92322, Ch\^{a}tillon, France}
\affiliation{European Theoretical Spectroscopy Facility (ETSF), B-4000 Sart Tilman, Li\`{e}ge, Belgium}
\author{Van Binh Vu}
\affiliation{Universit\'{e} Paris-Saclay, CEA, CNRS, SPEC, 91191 Gif-sur-Yvette, France}
\author{Elisa Serrano Richaud}
\affiliation{Universit\'{e} Paris-Saclay, ONERA, CNRS, Laboratoire d'\'{e}tude des microstructures (LEM), 92322, Ch\^{a}tillon, France}
\author{Hakim Amara}
\affiliation{Universit\'{e} Paris-Saclay, ONERA, CNRS, Laboratoire d'\'{e}tude des microstructures (LEM), 92322, Ch\^{a}tillon, France}
\affiliation{Universit\'{e} Paris Cit\'{e}, Laboratoire Mat\'{e}riaux et Ph\'{e}nom\`{e}nes Quantiques (MPQ), CNRS-UMR7162, 75013 Paris, France}
\author{Sylvain Latil}
\affiliation{Universit\'{e} Paris-Saclay, CEA, CNRS, SPEC, 91191 Gif-sur-Yvette, France}

\date{\today}

\begin{abstract}
We investigate the electronic structure and the optical absorption onset of close-to-30\degree twisted hexagonal boron nitride bilayers.
Our study is carried out with a purposely developed tight-binding model validated against DFT simulations. 
We demonstrate that approaching 30\degree (quasicrystal limit), all bilayers sharing the same moiré supercell develop identical band structures, irrespective of their stacking sequence.
This band structure features a bundle of flat bands laying slightly above the bottom conduction state which is responsible for an intense peak at the onset of the absorption spectrum.
These results suggest the presence of strong, stable and stacking-independent excitons in boron nitride 30\degree-twisted bilayers. 
By carefully analyzing the electronic structure and its spatial distribution, we elucidate the origin of these states as moiré-induced K-valley scattering due to interlayer B$-$B coupling. We take advantage of the the physical transparency of the tight-binding parameters to derive a simple triangular model based on the B sublattice that accurately describes the emergence of the bundle. 
Being our conclusions very general, we predict that a similar bundle should emerge in other close-to-30{\degree} bilayers, like transition metal dichalcogenides, shedding new light on the unique potential of 2D materials.
\end{abstract}


\maketitle



The unconventional physical properties exhibited by twisted bilayers have given rise to the field of twistronics~\cite{Carr2020,Liu2021, Andrei2021, Du2023}.
By stacking 2D atomic layers, a geometric moiré superlattice emerges as a lattice mismatch and/or a rotational twist~\cite{Gratias2023}.
The resulting pattern modulates the potential at the supercell scale and hence changes the electronic band structure typically through the formation of low dispersing bands often presenting peculiar properties~\cite{Yankowitz2019, Balents2020,Cao2018}.
Typical examples of moiré composites include the pioneering twisted bilayer graphene~\cite{Bistritzer2011}, twisted hexagonal boron nitride (hBN)~\cite{Xian2019, Liu2021, Kim2023} or hetero- and homobilayers of transition metal dichalcogenides (TMDs)~\cite{Scuri2020}.
In gapped twisted bilayers, like hBN, the width of band edge states decreases continuously with the angle of twist (no magic angle) and the presence of different atomic species generates several stacking possibilities with specific electronic properties, providing an additional degree of freedom with respect to graphene bilayers~\cite{Zhao2020, Walet2021, Latil2023}.

Since the early stages of twistronics, the scientific community has mostly focused on the small twist angle limit. 
In this context, continuous models~\cite{Scheer2022, Fang2015} or tight-binding (TB) hamiltonians~\cite{Trambly2010, Venkateswarlu2020, Long2022, Ochoa2020} have been developed and density functional theory (DFT) calculations~\cite{Trambly2010, Fang2015} carried out on several 2D material bilayers.
Actually, few works treat also larger twisting angles (in the 15\degree{}-28\degree range) ~\cite{Sboychakov2015,Pal2019, Mondal2023} and even less works consider rotations of 30° ~\cite{Ahn2018, Yao2018} or very close to it~\cite{Trambly2010, Moon2019}.
In any case, all these studies focus only on graphene bilayers and mention large angle configurations among other structures, none of them being specifically devoted to the investigation of close-to-30\degree{ }twists. Regarding hBN, only the 30\degree{ }twisted bilayers have been considered very recently~\cite{Chernozatonskii2023}. DFT calculations suggest that this BN material is a unique wide-gap 2D quasicrystal, yet its electronic and optical properties have never been addressed specifically.

In this Letter, we investigate the band structure and optical response of different stackings of hBN bilayer twisted in the vicinity of 30\degree{ } using a dedicated TB model.
We demonstrate that all structures develop an identical bundle of flat states just above the bottom of the conduction band.
We trace its physical origin and develop a simple model describing its formation.

To identify unambiguously the structures, we use the definitions introduced in~\cite{Latil2023} and recalled in Appendix A.
Our TB model is purposely designed to describe accurately the last occupied and the first empty states
Besides permitting calculations in supercells close enough to 30\degree, the TB formalism also allows to unravel some fundamental mechanism.
Our TB model is inspired by literature~\cite{Trambly2010, Trambly2012,Venkateswarlu2020} and detailed in Appendix B.
A feature it is worth stressing is the distance-dependent exponential decay of the interlayer hopping terms whose prefactor $\gamma^{XY}$ gives a measure of the interlayer coupling strength, $XY$ labelling the pairings $BN$, $BB$ and $NN$.
Remarkably, our TB model happens to describe reasonably well the top valence (TV) and bottom conduction (BC) states in a wide range of twist angles, comparing well to other low-angle models~\cite{Walet2021, Long2022, Henriques2022, Roman-Taboada2023}.

\begin{figure}
\centering
\includegraphics[width=\linewidth]{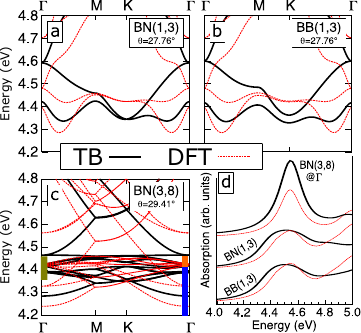}
\caption{$(a,b,c)$: Conduction bands of BN(1,3), BB(1,3) and BN(3,8) in TB (black solid) and DFT (red dashed). The TV of all structures have been aligned to 0.0~eV. In $(c)$, tick bars on the canvas highlight notable energy intervals (see text). $(d)$: Onset of IP absorption spectra of the same systems. The BN(3,8) spectra are computed only in the $\Gamma$ point. All spectra have been broadened with a Lorentzian with variance 0.1~eV.}
\label{fig:figure_validation_unique}
\end{figure}

The quality of the parametrization can be appreciated in Figure~\ref{fig:figure_validation_unique}  where we report the BC and independent-particle (IP) optical spectra computed both with our TB model and ab-initio methods.
Ab-initio simulation details are reported in Appendix C.
We present results from the BB(1,3), BN(1,3) and BN(3,8) bilayers, chosen as paradigmatic because of specific characteristics of their band structure, but we checked that the agreement is equally good in the other stackings (cfr. Appendix B and Figure~\ref{fig_sm:validation_bands_all_stackings}).

Our parametrization reproduces the general dispersion of DFT bands and the gapwidth.
Particular attention has been paid on the BC.
Let us first consider panels $(a)$ and $(b)$, i.e. the two (1,3) supercells.
DFT predicts the formation of a pretty flat dispersion in the M-K region in both systems. 
Actually the two bands avoid each other in the BB(1,3), even though the splitting is extremely small. Instead, they cross at K in the BN(1,3), consistently with what simulated at smaller angles~\cite{Latil2023}. 
Our TB model catches very well these features although the splitting in the BB stacking is somewhat overestimated.
At a larger angles, like in the BN(3,8), the agreement is even better as panel $(c)$ exemplifies well.
In particular, the model predicts correctly the emergence of a group of densely packed and low dispersing bands concentrated between 4.37~eV and 4.46~eV, that we will call the `bundle' of flat states, highlighted by an olive green side bar in Figure~\ref{fig:figure_validation_unique}c). 
It is useful to split the conduction bands into a lower energy region (blue bar in Figure~\ref{fig:figure_validation_unique}.c) called `shallow conduction', and a higher energy region where bands are particularly flat called the `deep bundle' region (orange bar in Figure~\ref{fig:figure_validation_unique}c).
All energy intervals are given with respect to the top of the valence band.


We also evaluated with the two methods the imaginary part of the IP dielectric function $\varepsilon(\omega)$ in the same three systems whose results are reported in panel $d$ of Figure~\ref{fig:figure_validation_unique}.
Details on the calculation can be found in Appendices B and C.
Because of the large size of the (3,8) supercell (388 atoms), the ab-initio spectrum is computed only in the $\Gamma$ point, and the same in TB for sake of comparison.
Both methods predict a well-detached peak at 4.5 eV, corresponding to transitions towards the bottom conduction states.
Differences between the stackings are negligible, indicating that not only the band structure but also the wavefunctions are remarkably similar in these systems.

\begin{figure*}
    \centering
    \includegraphics[width=\linewidth]{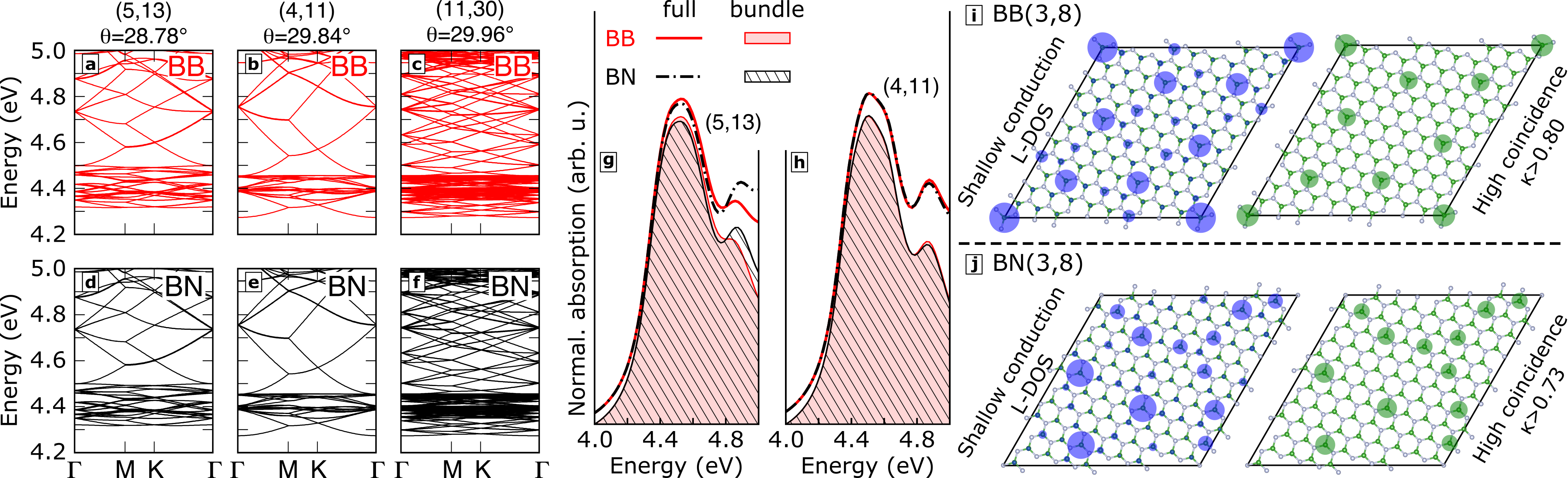}
    \caption{($a-c$): Conduction bands of the BB(5,13), BB(4,11) and BB(11,30) respectively. ($d-f$): Same as ($a-c$) in the BN stacking. The TV of all structures have been aligned to 0.0~eV. $(g, h)$: IP absorption spectra of the BB(5,13) and BB(4,11) (red solid curves) and BN(5,13) and BN(4,11) (black dashed) obtained including all empty and occupied states (full). Shaded and patterned areas are obtained by restricting the empty states between 4.34~eV and 4.48~eV. $(i,j)$: Blue circles: Radius proportional to the L-DOS in the shallow conduction interval. Green circles: Radius proportional to $\mathcal{C}_{j_B}(\mathbf{d})$ wherever it is higher than $\kappa$. See main text for details.
    Data are shown only for the upper layers of the BB(3,8) $(i)$ and BN(3,8) systems $(j)$. }
    \label{fig:figure_large-angles_bsp}
\end{figure*}

Having validated the TB model, we extend our investigation to twist angles closer to 30{\degree} and systems hardly attainable with DFT.
In panels $a$ to $f$ of Figure~\ref{fig:figure_large-angles_bsp}, we report the BC states of the BB and BN stackings in the (5,13), (4,11) and (11,30) supercells, corresponding to twist angles ranging from 28.78{\degree} to 29.96\degree.
The tendency observed already in the (1,3) and (3,8) supercells is here confirmed and strengthened: the stackings present basically the same band structure at fixed supercell.
This is actually true for all five stacking, as we assess in Figure~\ref{fig_sm:large_angle_conduction_all_stackings} and Appendix F.
The indistinguishability of the stacking sequence when approaching 30\degree{ }twist comes from the fact that, in this limit, the bilayer approaches a quasicrystal without any translation symmetry. 
As a consequence, all local configurations are realized somewhere and a sort of self-similarity arise such that in each supercell approximate replicas of smaller cells of all the five stackings can be found.
We will encounter another manifestation of this property later on.
Actually, the same happens in any homobilayer formed of hexagonal monolayers, so we expect a similar indistinguishability to occur also in close-to-30\degree{ }twisted graphene, TMDs, silicene and many of the most popular 2D materials.

More interestingly, we observe a bundle of flat states forming in the conduction band of all structures at all angles, comprised in a narrow interval (about 100~meV) centered around 4.40~eV.
This is in contrast to small-angle twisted hBN bilayers, where one or more single states are formed directly in the gap and clearly separated in energy by about 0.1 eV~\cite{Xian2019,Zhao2020,Liu2021}.
Because driven by the twist, the corresponding enhancement of the density of states (DOS) is very robust as basically independent on the stacking sequence or on the chemical environment. It is hence deterministic, contrary to a similar enhancement observed in bulk quasicrystals~\cite{Fujiwara90,Roche97,Krajci2001}.

Our TB formalism allows us to identify easily the origin of the bundle in terms of the monolayer valleys. In fact, since our TB model contain no \emph{s} orbitals, a $\Gamma$ component can be excluded from the beginning.
A deeper analysis, detailed in Appendix D, indicates that the bundle comes in fact from states close to the $K$ and $M$ points of the monolayer Brillouin zone.
This explains the emergence of similar bundles in graphene-based incommensurate stackings~\cite{Moon2019,Yu2019,Uri2023a} (even though in these systems they are created  symmetrically also in the valence) and suggests that similar bundles should emerge in other hexagonal 2D materials, namely TMDs.

It is particularly worthwhile to study the impact of these flat states on the absorption properties.
We used TB to compute
the IP optical response in the (5,13) and (4,11) supercells, here reported in Figure~\ref{fig:figure_large-angles_bsp}, panels $g$ and $h$. 
As expected, the BN and the BB stackings  present very small differences that are further washed out as the twist angle approaches 30\degree.
Spectral onsets are dominated by the same intense and well detached peak observed in Figure~\ref{fig:figure_validation_unique}.d).
Spectra including only transitions toward the bundle (4.34~eV to 4.48~eV) recover essentially the same signal.
This demonstrates that quasicrystallinity has a strong impact on the smallest excitations of this material because the bundle, despite not being the BC, is solely responsible of the absorption onset.
In this respect, BN-based quasicrystals are very different from graphene ones~\cite{Moon2019,Yu2019,Uri2023a} where the low-energy physics is still dominated by monolayer-like Dirac cones.
Given these results, we predict that close-to-30\degree{} twisted hBN bilayers will display exceptionally strong, robust (and possibly localised) electron-hole excitations.

To go beyond in the analysis, we look at the spatial distribution of the conduction states by evaluating  the local DOS (L-DOS) in energy intervals corresponding to the deep bundle and the shallow conduction highlighted in Figure~\ref{fig:figure_validation_unique}.c).
We show results in the upper layer of the (3,8) supercells, because pictures are easier to read than in larger structures.
In Figure~\ref{fig:figure_large-angles_bsp}, the radius of the blue circles in panels $i$ and $j$ is proportional to the L-DOS in the upper layer of the BB(3,8) and BN(3,8) respectively.
No DOS is centered on N sites since they contribute only to valence states~\cite{Galvani2016}.
Despite the resemblance of both band structures and optical spectra,
the two stacking develop quite different patterns
which actually hide fascinating similarities. In fact inside each structure, one can find infinite rearrangements of smaller-cell approximants of all the five stackings repeating themselves in a kind of self-similar scheme.
The L-DOS patterns of Figure~\ref{fig:figure_large-angles_bsp} arise from a sort of frustrated interference between these lower order configurations.
Examples of this are presented in Appendix G, but further studies go beyond the scope of this article.
The L-DOS appears to be stronger on sites where B atoms of the two layers are almost vertically aligned.
To highlight better this feature, we evaluate at each B site $j_B$ the coincidence function $\mathcal{C}_{j_B}(\mathbf{d}) = 1 - d_{xy}/D $, where $d_{xy}$ is the in-plane component of the vector $\mathbf{d}$ connecting the site $j_B$ of one layer to the closest B site of the other layer and $D=1.452$~\AA{ }is the in-plane interatomic distance.
By definition, $\mathcal{C}_{j_B}(\mathbf{d})=1$ if two $j_B$ and a B atom of the other layer are vertically aligned, and decreases linearly to 0 where $j_B$ is aligned with a N or a hexagon center.
The radius of green circles of Figure~\ref{fig:figure_large-angles_bsp} $i$ and $j$, is proportional to $\mathcal{C}_{j_B}(\mathbf{d})$ of all B sites of the upper layer for which $\mathcal{C}_{j_B}(\mathbf{d})> \kappa$ where $\kappa=0.8$ in BB(3,8) and $\kappa=0.73$ in BN(3,8).
The resemblance between the high-coincidence patterns and the shallow conduction L-DOS is striking.
The same can be shown for the ``deep bundle" states and low coincidence patterns (cfr. Appendix F and Figure~\ref{fig:coincidence-and-ldos_4-11}).
This analysis reveals that B$-$B interlayer states play a crucial role in the bundle formation and that the more vertical the B$-$B alignment, the stronger the coupling and hence the lower the energy of the corresponding empty state.
These are actually bonding states because the TB coefficients of coinciding and quasi-coinciding sites change sign in the two layers.

\begin{figure}[b]
    \centering
    \includegraphics[width=1.0\linewidth]{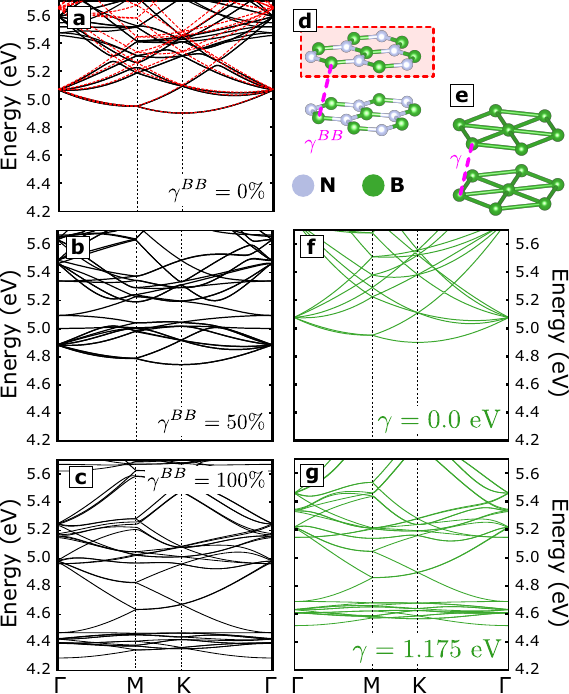}
    \caption{($a$): TB conduction band of the monolayer in the (3,8) supercell (dashed red) and of the BB(3,8) with $\gamma^{BB}=0$~eV. ($b$ and $c$): The same bilayer with $\gamma^{BB}$ equal to 50\% and 100\% of the correct value. The top valence of all structures have been aligned to 0.0 eV. ($d$): Ball and stick model of the actual twisted bilayer. Dashed magenta line: the $\gamma^{BB}$ interlayer coupling. Red shaded area: the isolated monolayer. ($e$) Ball and stick model of the triangular lattice twisted bilayer made only of B sites. Dashed magenta line: the $\gamma$ interlayer coupling. ($f$ and $g$): bands of the triangular model with $\gamma$= 0~eV and 1.715~eV respectively.}
    \label{fig:hopping_analysis}
\end{figure}

Inspired by this analysis, we calculate the band structure of the BN(3,8) including all TB parameters except for the prefactor of the B$-$B interlayer coupling $\gamma^{BB}$ which we set to zero.
The resulting states are reported in Figure~\ref{fig:hopping_analysis} a). They are dispersing and are basically indistinguishable from those of a monolayer.
If we now set  $\gamma^{BB}=1.225$~eV (50\%) (panel b), then localized states begin forming until a bundle of completely flat bands emerges at $\gamma^{BB}=2.45$~eV (100\%) in panel c. 
The weakness of the interlayer N$-$N coupling explains why there is no such a feature in the valence band, contrary to what happens in graphene twisted bilayers where conduction-conduction and valence-valence couplings are equivalent~\cite{Yu2023}.

We then devise an even simpler TB model which describes the formation of the bundle states. 
We select only the B sublattices obtaining a structure formed of two triangular lattice where only B$-$B interactions are taken into account.
Details of the model and its relation to the full honeycomb model are reported in Appendix E.
The conduction band structure of this simplified triangular model is reported in Figures~\ref{fig:hopping_analysis} f and g respectively for vanishing and non-vanishing interlayer coupling.
The model reproduces the isolated honeycomb monolayer at no coupling and gives rise to a bundle of flat bands at full coupling remarkably similar to the honeycomb TB model, clearly demonstrating that B$-$B interlayer interaction is solely responsible of the localization of the electrons in high-angle twisted BN bilayers.
This model allows us to unravel a fundamental mechanism common to other large-angle twisted bilayers.
We deem probable that a similar bundle of flat states will emerge  in the valence band of close-to-30\degree{ }twisted TMDs where the interlayer coupling is mostly due to chalcogen $p_{z}$ states which form the top of the valence band~\cite{Fang2015}.

To conclude, we have investigated the electronic and optical properties of hBN bilayers at twist angles close to 30\degree~by means of a purposely developed TB model. 
We have demonstrated that in this range of angles all hBN bilayers develop the same electronic properties, irrespective of the stacking sequence.
This is characterised by the emergence of a bundle of low-dispersing states right above the bottom of the conduction band. This results from a strong coupling between B atoms belonging to different layers and is responsible of an intense and robust peak at the onset of the absorption spectrum.
The asymmetry of the corresponding DOS enhancement and its dominant role in the absorption onset, make hBN quasicrystals very different from the other 2D quasicrystals studied so far (mainly 30°-twisted graphene bilayer~\cite{Moon2019,Yu2019}, and incommensurate stacks of graphene trilayers\cite{Uri2023a}.
We captured this fundamental mechanism with a very simple triangular lattice TB model which can be applied to many other twisted bilayers (e.g. homobilayers of TMDs).
Our results suggest that 30\degree -twisted BN bilayers may host extremely strong excitonic phenomena  originated by the bundle of flat bands and independent on the stacking sequence. Owing to its geometrical origin, similar bundles are expected to emerge in the valence band of TMDs quasicrystals, probaly with similar consequences on their excitonic properties.
Moreover, its robustness with respect to the stacking sequence is expected to be an ubiquitous characteristic in the quasicrystal limit, and to occur in twisted bilayers of other 2D materials including TMDs, antimonene, silicene, transition metal monochalcogenides, and all homostructures formed of hexagonal single-layers.

\begin{acknowledgments}
The authors acknowledge funding from the European Union’s Horizon 2020 research and innovation program under grand agreement N◦ 881603 (Graphene Flagship core 3) and from public grants overseen by the French National Research Agency (ANR) as part of the ‘Investissements d’Avenir’
program (Labex NanoSaclay, reference: ANR-10-LABX-0035) and under the EXCIPLINT project (Grant No. ANR-21-CE09-0016).
\end{acknowledgments}

\section{Appendix A: Structural definitions and nomenclature}
\label{sec:App_structure}
Here we report the structural details to reproduce our structures and recall the definitions and nomenclature\cite{Latil2023} most relevant for the current work.

The cell parameter of the honeycomb lattice is 2.54 \AA.
Every BN bilayer with hexagonal symmetry can be identified univoquely by a stacking label and a couple of integers $(q,p)$ defining the moiré supercell. 
These indeces define two matrices that, once applied to the monolayer unitary vectors, generate the supercell of the lower layer and the twisted supercell of the top layer.
Since the number of primitive cells required to span the supercell is
\begin{equation}
    \Omega = p^2+q^2+pq
\end{equation}
the total number of atoms in the bilayer supercell is $4\Omega$.

\begin{figure}
    \centering
    \includegraphics[width=\linewidth]{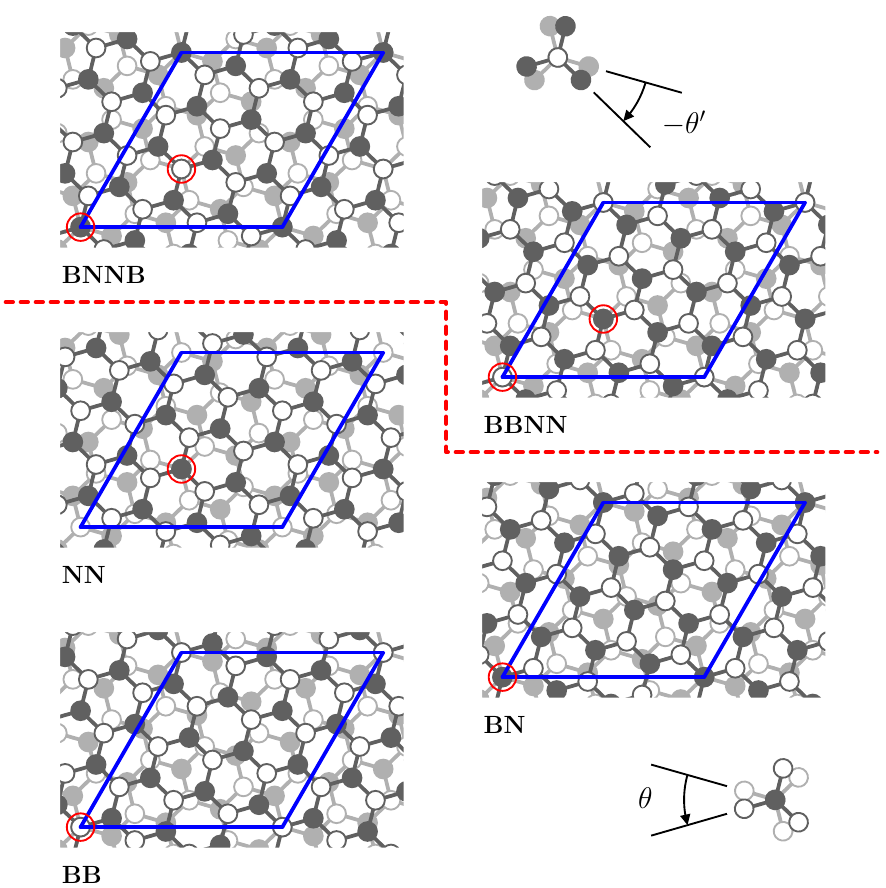}
    \caption{The five hexagonal stackings in the (1,3) moiré supercell. Red circles highlight the coincidence sites. A red dashed line separates the double-coincidence stackings with twist angle $-\theta'$ (top section) from the single-coincidence once with twist angle $\theta$ (bottom section).}
    \label{fig:structures_q1p3}
\end{figure}

Moreover, there exist 
only five hexagonal stacking sequences for each $(q,p)$-pair.
The corresponding five stacking labels take their name from the pair of atoms placed exactly on top of each other on the three high-symmetry points of the hexagonal supercell.
These are the origin (0~,~0), the point (1/3~,~1/3) and the point (2/3~,~2/3). 
The five stackings divide into single-coincidence stackings (labelled BB, BN and NN) and double coincidence stackings (labelled BBNN and BNNB).
As an example, we report in Figure~\ref{fig:structures_q1p3} all five stackings of the (1,3) supercell. 
If we take, without loss of generality, $p>q$ then the rotation of the top layer with respect to the bottom layer will be given either by an angle $\theta$ (in the single-coincidence systems, or ``atom-on-hexagon" geometries) or $-\theta'$ (in the double-coincidence case, or ``hexagon-on-hexagon" geometries) with $\theta'=\pi/3 - \theta$.
Actually, both twist angles are derived from $p$ and $q$ with specific formulae\cite{Latil2023}, namely:
\begin{equation}
    \begin{split}
    \tan\theta&=\sqrt{3}\frac{p^2-q^2}{p^2+q^2+4pq} \\
    \tan\theta'&=\sqrt{3}\frac{q^2+2pq}{2p^2-q^2+2pq}
    \end{split}
    \label{SM:equ:angles}
\end{equation}
Our work sheds light on supercells with angles in the vicinity of 30\degree. Therefore, from  
(\ref{SM:equ:angles}), we have to 
choose integer $(q,p)$-pairs that approximate
\begin{equation}
    \label{SM:equ:close30}
    p\simeq q(1+\sqrt{3})
\end{equation}
so that $\theta$ and $\theta'$ tend to 30{\degree} asymptotically. 
The best set of approximants of equation (\ref{SM:equ:close30}) are listed in Table \ref{SM:tab:close30}.

\begin{table}
    \centering
    \begin{tabular}{cccc}
        \hline \hline
        $(q,p)$ & $N_{\text{atoms}}$ & $\theta$ & $\theta'$ \\
        \hline
        \bf{(1,3)}   & \bf{52}   & \bf{32.20\degree} & \bf{27.80\degree} \\
        \bf{(5,13)} & \bf{1036} & \bf{28.78\degree} & \bf{31.22\degree} \\
        (11,28) & 4852 & 28.25\degree & 31.75\degree \\
        (6,17)  & 1708 & 30.87\degree & 29.13\degree \\
        \bf{(3,8)}   & \bf{388}  & \bf{29.41\degree} & \bf{30.59\degree} \\
        (9,25)  & 3724 & 30.40\degree & 29.60\degree \\ 
        \bf{(4,11)}  & \bf{724}  & \bf{30.16\degree} & \bf{29.84\degree} \\
        \bf{(11,30)} & \bf{5404} & \bf{29.96\degree} & \bf{30.04\degree} \\
        \hline \hline
    \end{tabular}
    \caption{The set of the best $(q,p)$-pairs that approximate equation (\ref{SM:equ:close30}), ordered from top to bottom by decreasing distance to 30\degree. The list is limited  to structures
    containing 5404 atoms.
    Bold: systems discussed in the main text.}
    \label{SM:tab:close30}
\end{table}



\section{Appendix B: Tight-Binding model}
\label{sec:App_tbmodel}

Since the hexagonal BN monolayer band structure is easily obtained with a first neighbors tight-binding Hamiltonian,
our basis is constituted of the $p_z$ orbitals of B and N.
The intralayer part of the Hamiltonian relies on three parameters: the on-site energies 
and the first-neighbor in-plane hopping 
\[
\varepsilon_B=4.90\text{~eV} \,,\quad \varepsilon_N=0.00\text{~eV} \,,\quad t_\parallel=-2.65\text{~eV.}
\]
Our rule for defining the interlayer hopping integrals is strongly inspired by a model originally developed for 
graphene moiré bilayers~\cite{Trambly2010, Trambly2012} and successively extended to twisted bilayer MoS$_2$~\cite{Venkateswarlu2020}. In this model, the matrix elements between two 
$p_z$ orbitals separated by vector $\mathbf{d}$ 
obey the Slater-Koster angular dependence 
like 
\[
t_\perp(\mathbf{d})=n^2 V_{pp\sigma}(d) + (1-n^2) V_{pp\pi}(d)\,,
\]
where $n=d_z/d$ is the vertical direction cosine.
The $V_{pp\sigma}$ and $V_{pp\pi}$ bond integrals follow an exponential form.
Differently to the model cited above ~\cite{Trambly2010} we firstly restrict this rule to interlayer matrix elements only, hence between
$p_z$ orbitals belonging to different layers.
Secondly, we include only the $\sigma$ component in the intelayer hopping which has finally the form 
\begin{equation}
\label{SM:equ:tperp}
t_\perp^{XY}(\mathbf{d})=n^2 \gamma^{XY} F^{XY}_c(d)  \exp\left[Q_{XY}(a_\perp-d)\right]
\end{equation}
with $a_\perp$ being the interlayer distance~\cite{Latil2023} 
\[a_\perp=3.22 \text{~\AA}\]
and $XY$ labelling the pairings $BN$, $BB$ or $NN$. 
The values of the $\gamma^{XY}$ and $Q_{XY}$ parameters are reported in Table~\ref{tab:tb_parameters}.
In (\ref{SM:equ:tperp}), the function $F^{XY}_c$ is the smooth cutoff function defined in reference 
[\onlinecite{Trambly2010}] as
\begin{equation}
    F^{XY}_c(d)=\left\{
    1+\exp\left[ (d-r^{XY}_c)/l_c\right]
    \right\}^{-1}\,,
\end{equation}
where $l_c=0.265$~{\AA}  and $r^{XY}_c$ is the selected cutoff, that depends on the value of $Q_{XY}$ according to the relation
\[
r^{XY}_c=a_\perp+\frac{\text{ln}(10^3)}{Q_{XY}}\,.
\]

\begin{figure*}[p]
\includegraphics[width=\textwidth]{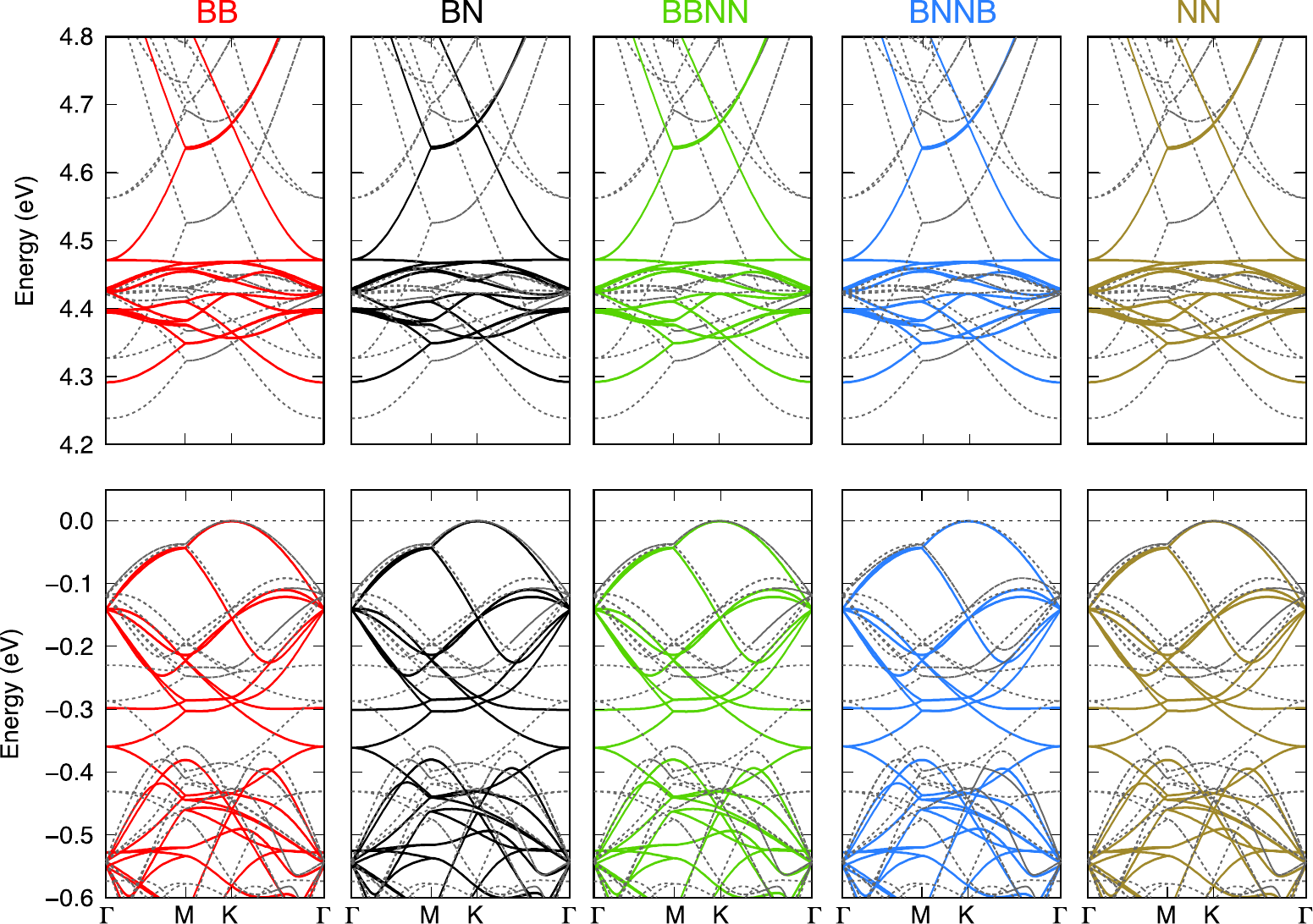}
\caption{Top valence (bottom panels) and bottom conduction states (top panels) of the (3,8) supercell in the BNNB, BN, NN, BB and BBNN stackings from left to right. Colored solid curves are from our tight-binding model, dotted grey curves from DFT.}
\label{fig_sm:validation_bands_all_stackings}
\end{figure*}

The performance of this model on the top valence and bottom conduction bands of all the five stackings in the (3,8) supercell can be appreciated in Figure~\ref{fig_sm:validation_bands_all_stackings}.

\begin{table}
\centering
    \begin{tabular}{cc | cc}
        \hline \hline
         prefactor          & value (eV) & decay & value (\AA$^{-1}$) \\
         \hline
         $\gamma^{BB}$      & 2.45  & $Q_{BB}$           & 3.0  \\
         $\gamma^{BN}$      & 0.75  & $Q_{BN}$           & 2.0  \\
         $\gamma^{NN}$      & 0.32  & $Q_{NN}$           & 1.6 \\
         \hline \hline
    \end{tabular}
    \caption{Set of parameters for  the interlayer hopping.}
    \label{tab:tb_parameters}
\end{table}

The imaginary part of the independent-particle transverse dielectric function $\varepsilon(\omega)$ is obtained at first order expansion in the coupling with the vector potential $\textbf{A}(\textbf{r},t)$ through the formula~\cite{GROSSO}:
\begin{equation}
  \varepsilon (\omega)=
  \frac{e^2 \pi}{m_0^2 \varepsilon_0 \omega^2} 
  \sum_{\textbf{k},m,\mu}
  \left\lvert \text{v}_{\kk\mu m}(\hat{\bm{e}}) \right\rvert^2
  \delta(E_{\kk\mu}-E_{\kk m}-\hbar\omega)\, ,
\label{eq:optics}    
\end{equation}
where $e$ is the 
elementary
charge, $m_0$ is the electron mass and $\varepsilon_0$ is the vacuum permittivity. 
In actual implementations, the delta function appearing in~\eqref{eq:optics} is actually replaced by a Lorentzian distribution the width of which has been fixed to 0.1~eV.
Expression~\eqref{eq:optics} describes the absorption of a
 photon with energy $\hbar\omega$ and polarization
vector $\hat{\textbf{e}}$ and the resulting promotion of an electron from the valence state $\ket{m,\textbf{k}}$ of energy $E_{\kk m}$ to the conduction
state $\ket{\mu,\textbf{k}}$ of energy $E_{\kk\mu}$.
The velocity matrix element
$\text{v}_{\kk\mu m}=\bra{\mu,\textbf{k}}\hat{\textbf{e}}\cdot\hat{\textbf{v}}\ket{m,\textbf{k}}$
is obtained from the eigenstates of the 
operator
$\hbar\hat{\textbf{v}}=i [\hat{\textbf{H}},\hat{\textbf{r}}]$.



\section{Appendix C: \textit{Ab-initio} calculation details}
\label{sec:App_abinitio}
The DFT simulations have been carried out with the free simulation package Quantum ESPRESSO~\cite{Giannozzi2009,Giannozzi2017}.
We used norm conserving pseudopotentials, a cutoff energy of 60.0~Ry for the wavefunctions and of 240~Ry for the charge. 
The exchange-correlation potential has been approximated with the Perdew-Burke-Ernzerhof model~\cite{Perdew1996}.
The Brillouin zone has been sampled with shifted Monkhorst-Pack~\cite{Pack1977} grids of $5\times5$ k-points in the $x$ $y$ plane in the (1,3) supercells and $3\times3$ in the (3,8) supercell.

\textit{Ab-initio} independent-particle absorption spectra have been calculated using the free simulation package Yambo~\cite{Marini2009,Sangalli2019}.
We sampled the Brillouin zone of the (1,3) supercells with a shifted Monkhorst-Pack~\cite{Pack1977} grid of $4\times4$ k-points and included 400 bands in the sum over states.
Note that this value of bands is actually much higher than what is required for the absorption onset alone.
In the BB(3,8) calculation, because of the larger size of the calculation which comprises 1552 electrons, we included bands with index ranging between 700 and 800 and we computed the sum over states only in the $\Gamma$ point.
In both cases we truncated the Coulomb interaction in the $z$ axis using the analytic formulation as implemented in Yambo~\cite{Marini2009,Sangalli2019}.

\section{Appendix D: Origin of the bundle as due to monolayer valley scattering}
\label{sec:App_monolayer_valleys}

\begin{figure*}
    \includegraphics[width=0.9\linewidth]{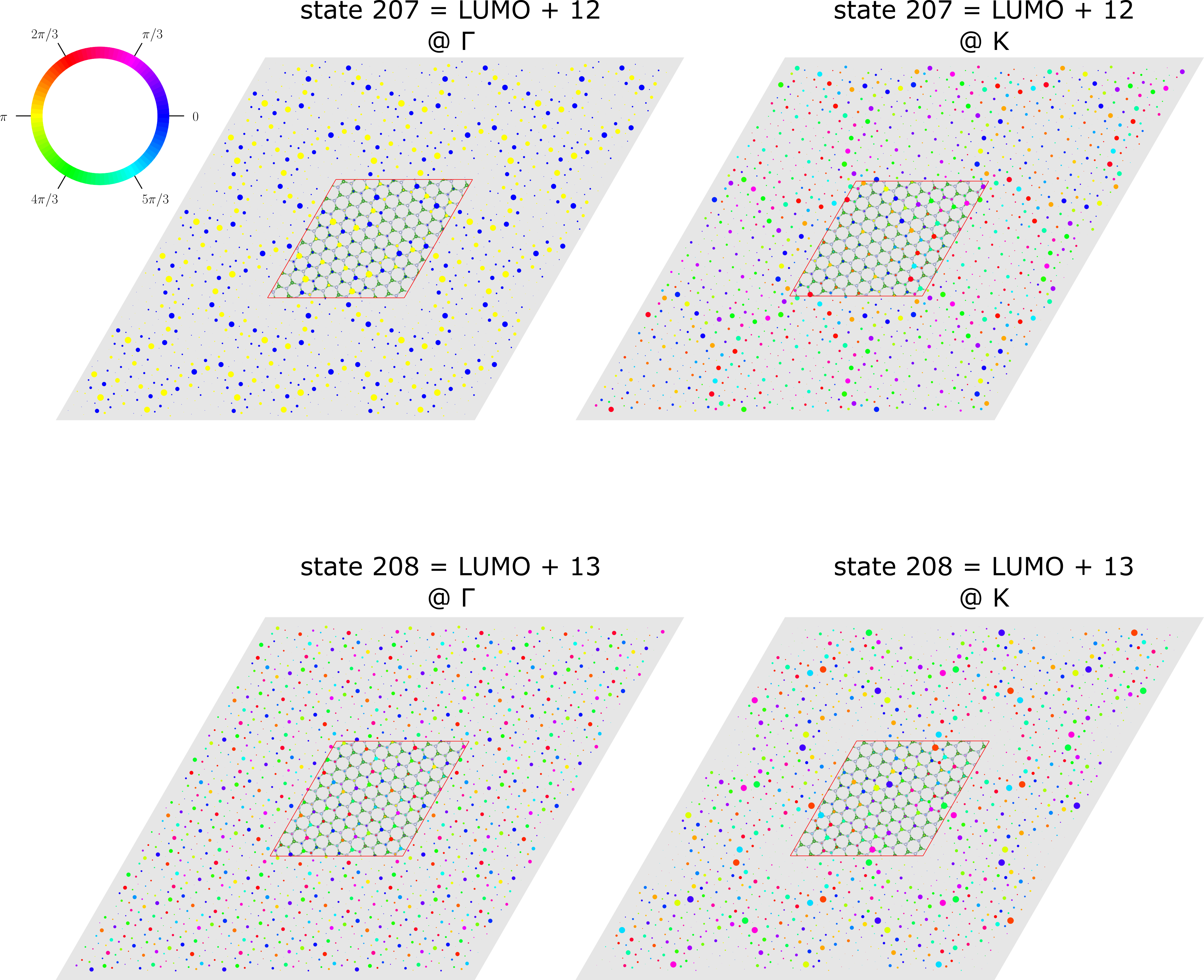}
    \caption{The TB complex coefficients of the eigenvectors of states 207 and 208 of the BNNB(3,8) at points $\Gamma$ (left) and K (right) of the moiré supercell. The bottom conduction is state number 195. Size of circles: modulus of the coefficient. Color of circles: complex phase (color code on top left corner). Only coefficients of the bottom layer are reported. In the centre of each plot, the structure of the bottom layer in the moiré supercell are also reported.}
    \label{fig:fluctuations}
\end{figure*}

\begin{figure}
    \includegraphics[width=\linewidth]{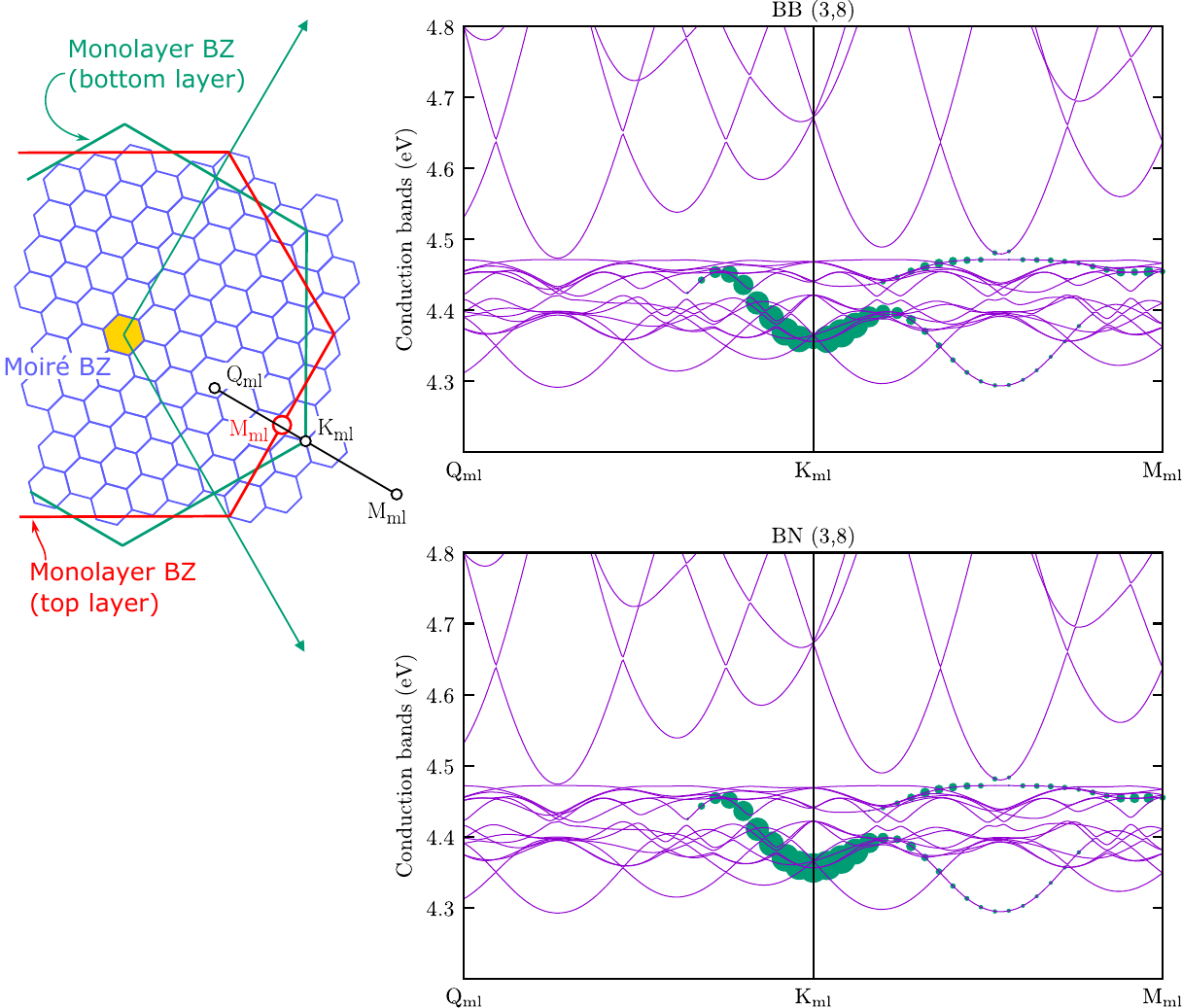}
    \caption{ Purple solid line: The band structure of the BB(3,8) and BN(3,8) plotted across several moiré Brillouin zones along the line which connects the points Q$_{\text{ml}}$, K$_{\text{ml}}$ and M$_{\text{ml}}$ of the hBN monolayer (Q$_{\text{ml}}$ being the point half-way between $\Gamma$ and K$_{\text{ml}}$). Green circles: Result of the projection of the bilayer states on the states of a hBN single layer oriented as the bottom layer. The size of the circle is proportional to the projection, i.e. the larger is a circle, the higher is the resemblance between the bilayer eigenstate and a monolayer Bloch state.
    }
    \label{fig:unfolding}
\end{figure}

In monolayer hBN, the bottom conduction at $\Gamma$ and K, even though very close in energy (see e.g. \cite{Galvani2016}), have very distinct orbital character: at $\Gamma$ they are pure $\sigma$ states whereas at $K$ (and $M$) they are pure
$\pi$ states.
The basis of our TB model is formed only of $p_z$ orbitals which explains the absence of the $\Gamma$ valley in Figure~\ref{fig:SM:figureML}.
As a consequence, in our model the bundle of flat bands can not have contributions from the monolayer $\Gamma$ valley.
On the other hand, the very good agreement with DFT simulations, makes us conclude that the $p_z$ basis is sufficient and appropriate for describing the mechanism of the bundle formation.
So, we claim that $\sigma$ states are definitely not involved in the formation of the bundle states.
As a further confirmation, we report in Figure~\ref{fig:fluctuations} an analysis of the phase of the TB eigenvectors.
In the plot, we report the TB complex coefficients of the eigenvectors of some states belonging to the bundle of the BNNB(3,8) structure in the points $\Gamma$ (left) and K (right) of the moiré supercell.
The size of each circle is proportional to the modulus of the coefficient and its colour to the phase according to the key on the top left corner.
The analysis demonstrates that the bundle states display a rapid phase modulation not compatible with a small momentum in the monolayer Brillouin zone, hence excluding a $\Gamma$ component.
To go even deeper in the analysis, we applied the unfolding technique described by Allen and coworkers\cite{Allen2013} along a straight path in the bottom layer Brillouin zone going from a point $Q_{\text{ml}}$ at half-way between $\Gamma_{\text{ml}}$ and $K_{\text{ml}}$ and the point $M_{\text{ml}}$ (the subscript $\text{ml}$ indicate that the frame of reference is the monolayer Brillouin zone).
The path passes then through the $K_{\text{ml}}$ point of the monolayer.
As displayed in Figure~\ref{fig:unfolding}, we found that the origin of the bundle is related to the moiré-induced scattering of states close to $K_\text{ml}$ and mostly laying along the $K_\text{ml}-M_\text{ml}$ line.

\section{Appendix E: Details of the triangular model}
\label{sec:App_triangular}

\begin{figure}
\centering
\includegraphics[width=\linewidth]{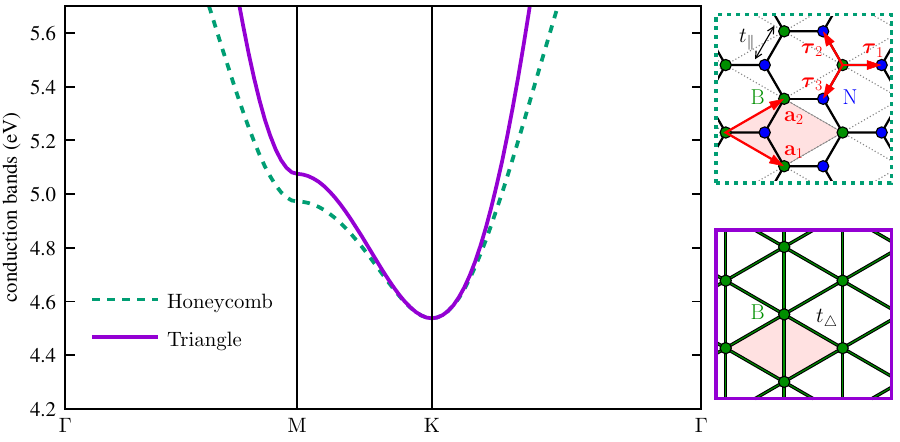}
\caption{The triangular model used to approximate the conduction bands in the Figure 4 of the main manuscript. }
\label{fig:SM:figureML}
\end{figure}

In our notation, the monolayer conduction band within the BN primitive cell writes
\begin{equation}
    \begin{split}
    E_c(\mathbf{k})
    &=
    \varepsilon_B^2/2+  
    \sqrt{
    \varepsilon_B^2/4+
    t_{\parallel}^2\left\lvert f(\mathbf{k})\right\rvert^2
    }
    \\
    f(\mathbf{k})
    &=
    \sum_{j=1,2,3} 
    \exp\left(
    i\mathbf{k}\cdot\boldsymbol{\tau}_j
    \right) \,.
    \end{split}
\end{equation}
As shown in a previous work~\cite{Galvani2016}, in the vicinity of the gap, the conduction eigenstates are
mainly localised on boron sites, which is related to the first term expansion
\begin{equation}
\label{equ:SM:conduc}
     E_c(\mathbf{k})
     \simeq
     \varepsilon_B+\frac{t_{\parallel}^2}{\varepsilon_B}
     \left\lvert f(\mathbf{k})\right\rvert^2\,.
\end{equation}
Actually, expression~\eqref{equ:SM:conduc} is identical to the unique eigenvalue of a first-nearest-neighbour TB model made on the triangular lattice formed by the boron sites.
The B$-$B hopping integral and the on-site energy of such a triangular lattice model can be replaced respectively by 
\begin{equation}\begin{split}
    t_{\bigtriangleup}
    &=
    \frac{t_{\parallel}^2}{\varepsilon_B} \quad \text{and}
    \\
    E_{\bigtriangleup}
    &=
    \varepsilon_B+3t_{\bigtriangleup}\,.
\end{split}\end{equation}
Finally, the interlayer hopping integrals (between B atoms only) follow the same construction rule as that for the honeycomb calculations, defined in 
(\ref{SM:equ:tperp}). The numerical value of $\gamma=1.175$~eV indicated in the main manuscript has been set by comparison with the twisted honeycomb TB model (cfr. Figure~\ref{fig:SM:figureML}).

\section{Appendix F: Band structure and local density of states of all stackings approaching 30\degree}
\label{sec:App_allbands}
\begin{figure}
    \centering
    \includegraphics[width=0.9\linewidth]{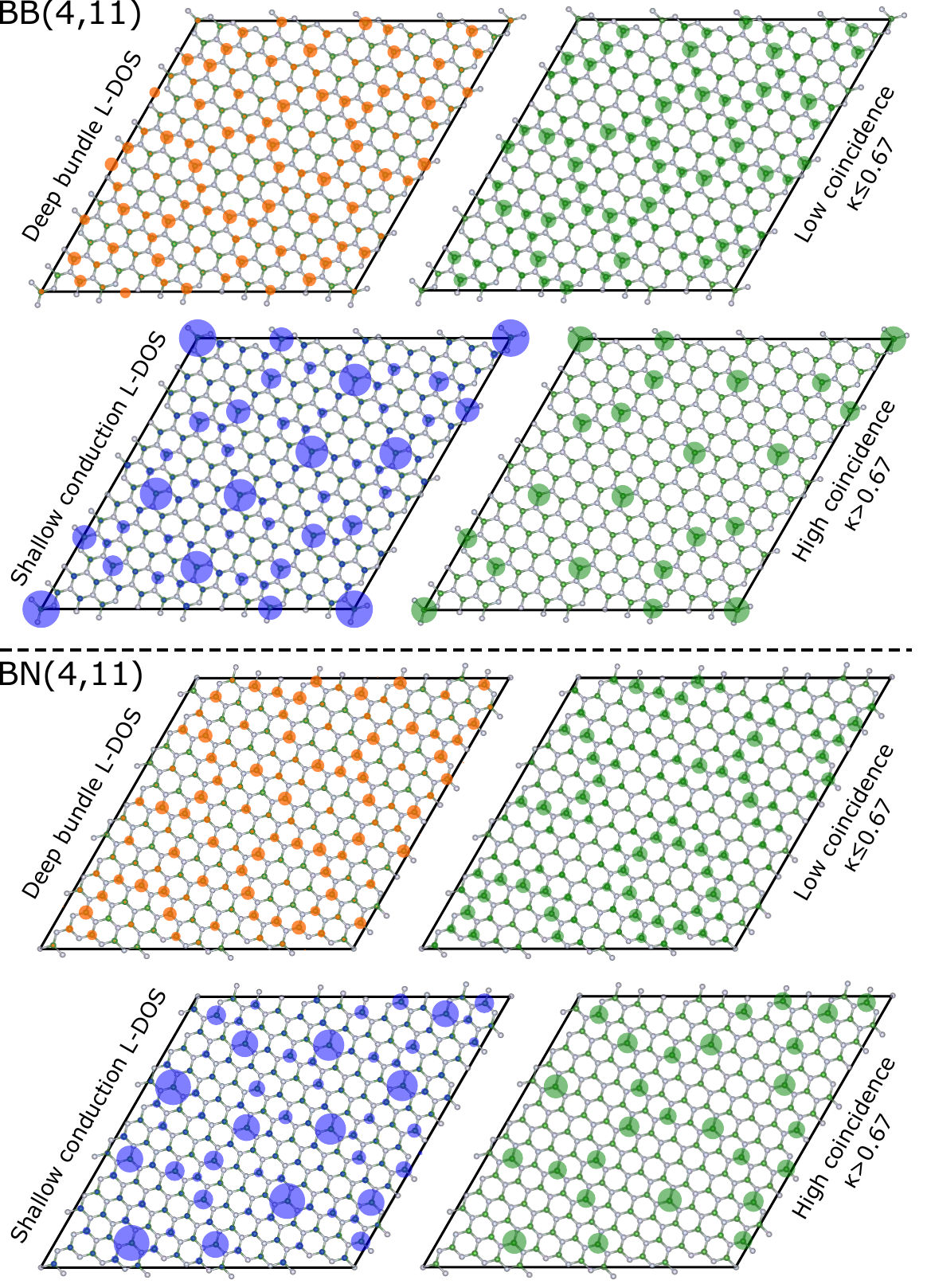}
    \caption{Orange circles: Radius proportional to the L-DOS in the deep bundle energy interval. Blue circles: As orange ones but in the shallow conduction interval. Green circles: Radius proportional to $\mathcal{C}_{j_B}(\mathbf{d})$ wherever it is lower than $\kappa$ or higher. See main text for details. 
    Data are shown only for the upper layers of the BB(4,11) and BN(4,11) systems.
    }
    \label{fig:coincidence-and-ldos_4-11}
\end{figure}

\begin{figure*}[p]
\includegraphics[width=\textwidth]{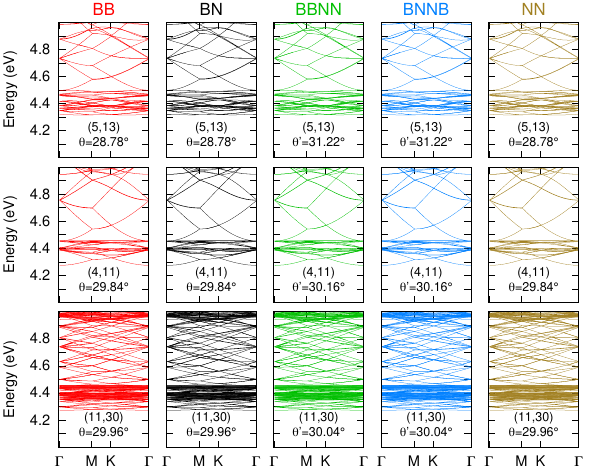}
\caption{From left to right: The tight-binding conduction bands of the BB, the BN, the BBNN, the BNNB and the NN stacking sequences. From top to bottom: Supercells (5,13), (4,11) and (11,30) with twist angles approaching 30\degree. The twist angle is either $\theta$ in single-coincidence stackings or $-\theta'$ in double coincidence ones.}
\label{fig_sm:large_angle_conduction_all_stackings} 
\end{figure*}

In this section we provide additional TB results on the large angle limit. 
In Figure~\ref{fig_sm:large_angle_conduction_all_stackings} we present the TB conduction bands of all five stackings in the (5,13), the (4,11) and the (11,30) supercells. 
In Figure~\ref{fig:coincidence-and-ldos_4-11}
we report the L-DOS and the coincidence map in the upper layers of the BB(4,11) and the BN(4,11) bilayers. 
Like in the main text, we divide the L-DOS the density of the `shallow conduction' states (blue circles) whose energies lie below 4.19~eV, and the `deep bundle' states (orange circles) with energies comprised between  4.19~eV and 4.46~eV, like in the (3,8) supercells.
Analogously, the coincidence map is split into high coincidence sites and low coincidence, the separation being fixed at $\kappa=0.67$.


\section{Appendix G: About the self-similar repeating patterns}
\label{sec:App_selfsimilar}
\begin{figure*}[p]
\includegraphics[width=0.9\textwidth]{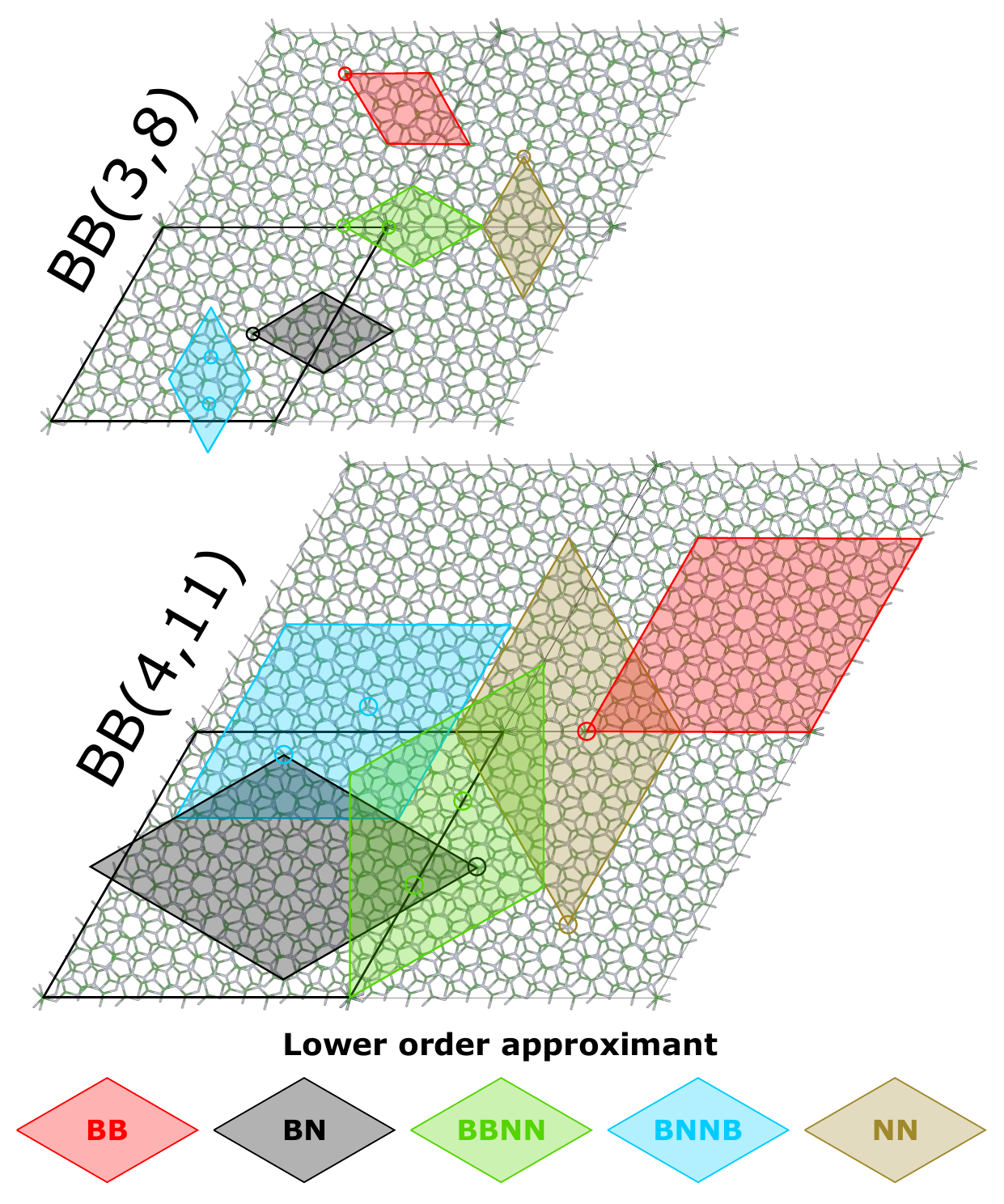}
\caption{Four supercells of the BB(3,8) (top) and the BB(4,11) (bottom) bilayers where both layers are visible. In each structure, highlighted in different colours are some local replicas of lower-order approximants of all the five stackings. Small colored circles highlight the corresponding coincidence site of each lowe-order approximant.}
\label{fig_sm:BB_self-similar} 
\end{figure*}

\begin{figure*}[p]
\includegraphics[width=0.9\textwidth]{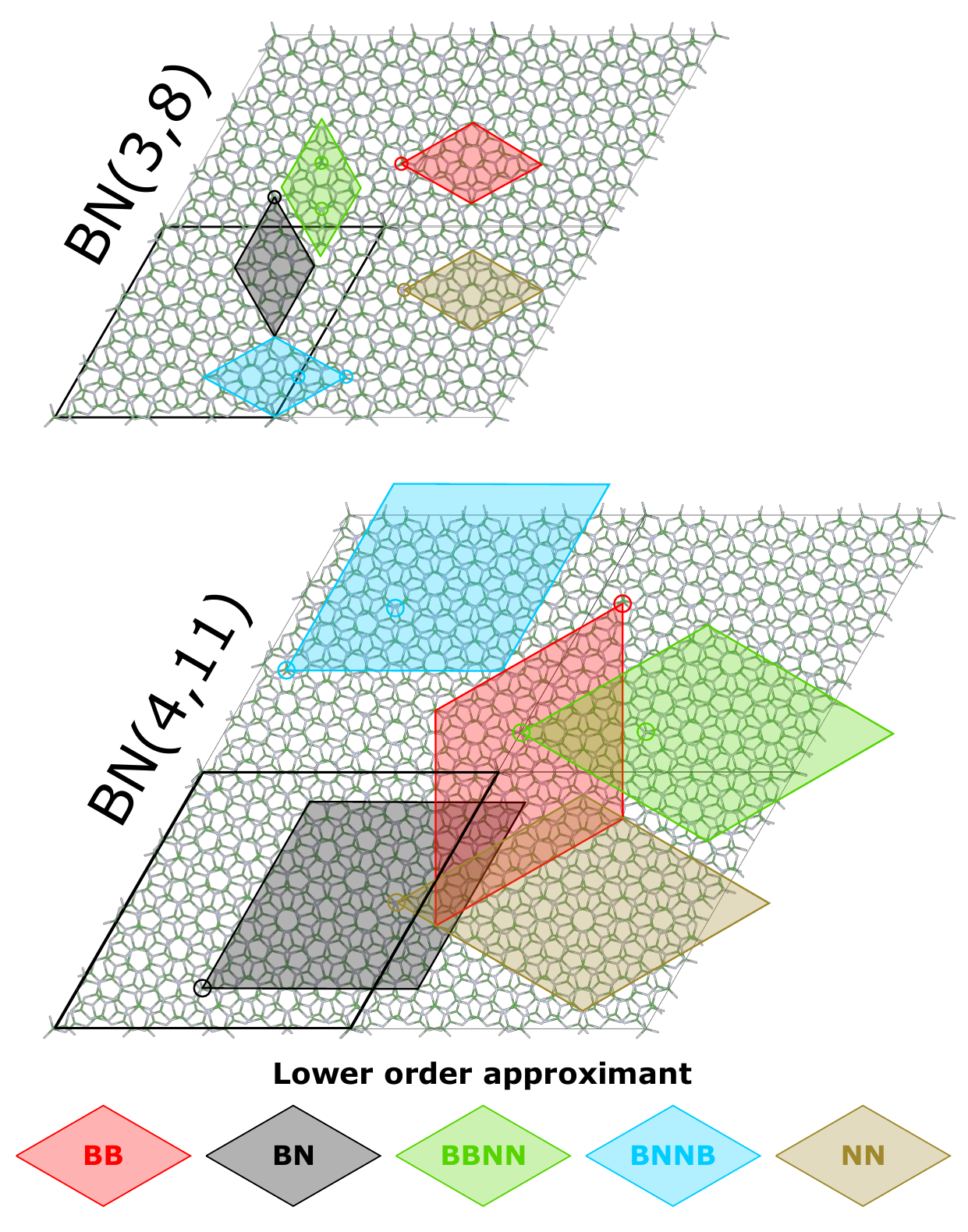}
\caption{Four supercells of the BN(3,8) (top) and the BN(4,11) (bottom) bilayers where both layers are visible. In each structure, highlighted in different colours are some local replicas of lower-order approximants of all the five stackings. Small colored circles highlight the corresponding coincidence site of each lowe-order approximant.} 
\label{fig_sm:BN_self-similar} 
\end{figure*}

In this section we illustrate with some images the presence of self-similar patterns repeating inside each bilayer.
We show that the moiré supercell of the approximant of a given stacking contains the lower-order approximants of all the stackings.
In effect, by construction one given approximant can not be obtained as a simple repetition of lower-size approximants.
As a consequence, it is impossible to tile perfectly a given approximant with a lower-order one.
So the inclusion and the repetition of the smaller cells within a bigger one is not perfect, which gives rises to a sort of frustration. 
The interference resulting from the superposition of this frustrated self-similar repetition of the lower-order approximants of all stackings is at the origin of the L-DOS patterns of the BB(3,8) and BN(3,8) reported in the main text, and of the BB(4,11) and BN(4,11) reported here in Figure~\ref{fig:coincidence-and-ldos_4-11}
As an illustration of this intriguing phenomenon, we report in Figure~\ref{fig_sm:BB_self-similar} four supercells of the BB(3,8) and the BB(4,11) bilayers. In the former (top panel), we highlight with different colours local configurations that are similar to the five stackings of the (1,3) supercell.
In the latter (bottom panel), we highlight local replicas of the five stackings of the (3,8) supercell using the same color code.
To help the identification, we added circles in the regions where the characteristic coincidence is realised. 
The scheme can be repeated again and again at all scales.
We give the same kind of representation in 
Figure~\ref{fig_sm:BN_self-similar}
for the BN stacking sequence.

\begin{thebibliography}{47}%
\makeatletter
\providecommand \@ifxundefined [1]{%
 \@ifx{#1\undefined}
}%
\providecommand \@ifnum [1]{%
 \ifnum #1\expandafter \@firstoftwo
 \else \expandafter \@secondoftwo
 \fi
}%
\providecommand \@ifx [1]{%
 \ifx #1\expandafter \@firstoftwo
 \else \expandafter \@secondoftwo
 \fi
}%
\providecommand \natexlab [1]{#1}%
\providecommand \enquote  [1]{``#1''}%
\providecommand \bibnamefont  [1]{#1}%
\providecommand \bibfnamefont [1]{#1}%
\providecommand \citenamefont [1]{#1}%
\providecommand \href@noop [0]{\@secondoftwo}%
\providecommand \href [0]{\begingroup \@sanitize@url \@href}%
\providecommand \@href[1]{\@@startlink{#1}\@@href}%
\providecommand \@@href[1]{\endgroup#1\@@endlink}%
\providecommand \@sanitize@url [0]{\catcode `\\12\catcode `\$12\catcode
  `\&12\catcode `\#12\catcode `\^12\catcode `\_12\catcode `\%12\relax}%
\providecommand \@@startlink[1]{}%
\providecommand \@@endlink[0]{}%
\providecommand \url  [0]{\begingroup\@sanitize@url \@url }%
\providecommand \@url [1]{\endgroup\@href {#1}{\urlprefix }}%
\providecommand \urlprefix  [0]{URL }%
\providecommand \Eprint [0]{\href }%
\providecommand \doibase [0]{http://dx.doi.org/}%
\providecommand \selectlanguage [0]{\@gobble}%
\providecommand \bibinfo  [0]{\@secondoftwo}%
\providecommand \bibfield  [0]{\@secondoftwo}%
\providecommand \translation [1]{[#1]}%
\providecommand \BibitemOpen [0]{}%
\providecommand \bibitemStop [0]{}%
\providecommand \bibitemNoStop [0]{.\EOS\space}%
\providecommand \EOS [0]{\spacefactor3000\relax}%
\providecommand \BibitemShut  [1]{\csname bibitem#1\endcsname}%
\let\auto@bib@innerbib\@empty
\bibitem [{\citenamefont {Carr}\ \emph {et~al.}(2020)\citenamefont {Carr},
  \citenamefont {Fang},\ and\ \citenamefont {Kaxiras}}]{Carr2020}%
  \BibitemOpen
  \bibfield  {author} {\bibinfo {author} {\bibfnamefont {S.}~\bibnamefont
  {Carr}}, \bibinfo {author} {\bibfnamefont {S.}~\bibnamefont {Fang}}, \ and\
  \bibinfo {author} {\bibfnamefont {E.}~\bibnamefont {Kaxiras}},\ }\bibfield
  {title} {\enquote {\bibinfo {title} {{Electronic-structure methods for
  twisted moir{\'{e}} layers}},}\ }\href {\doibase 10.1038/s41578-020-0214-0}
  {\bibfield  {journal} {\bibinfo  {journal} {Nat. Rev. Mater.}\ }\textbf
  {\bibinfo {volume} {5}},\ \bibinfo {pages} {748--763} (\bibinfo {year}
  {2020})}\BibitemShut {NoStop}%
\bibitem [{\citenamefont {Liu}\ \emph {et~al.}(2021)\citenamefont {Liu},
  \citenamefont {Xian}, \citenamefont {Mu}, \citenamefont {Zhao}, \citenamefont
  {Liu}, \citenamefont {Rubio},\ and\ \citenamefont {Wang}}]{Liu2021}%
  \BibitemOpen
  \bibfield  {author} {\bibinfo {author} {\bibfnamefont {B.}~\bibnamefont
  {Liu}}, \bibinfo {author} {\bibfnamefont {L.}~\bibnamefont {Xian}}, \bibinfo
  {author} {\bibfnamefont {H.}~\bibnamefont {Mu}}, \bibinfo {author}
  {\bibfnamefont {G.}~\bibnamefont {Zhao}}, \bibinfo {author} {\bibfnamefont
  {Z.}~\bibnamefont {Liu}}, \bibinfo {author} {\bibfnamefont {A.}~\bibnamefont
  {Rubio}}, \ and\ \bibinfo {author} {\bibfnamefont {Z.~F.}\ \bibnamefont
  {Wang}},\ }\bibfield  {title} {\enquote {\bibinfo {title} {{Higher-Order Band
  Topology in Twisted Moir{\'{e}} Superlattice}},}\ }\href {\doibase
  10.1103/PhysRevLett.126.066401} {\bibfield  {journal} {\bibinfo  {journal}
  {Phys. Rev. Lett.}\ }\textbf {\bibinfo {volume} {126}},\ \bibinfo {pages}
  {066401} (\bibinfo {year} {2021})}\BibitemShut {NoStop}%
\bibitem [{\citenamefont {Andrei}\ \emph {et~al.}(2021)\citenamefont {Andrei},
  \citenamefont {Efetov}, \citenamefont {Jarillo-Herrero}, \citenamefont
  {MacDonald}, \citenamefont {Mak}, \citenamefont {Senthil}, \citenamefont
  {Tutuc}, \citenamefont {Yazdani},\ and\ \citenamefont {Young}}]{Andrei2021}%
  \BibitemOpen
  \bibfield  {author} {\bibinfo {author} {\bibfnamefont {E.~Y.}\ \bibnamefont
  {Andrei}}, \bibinfo {author} {\bibfnamefont {D.~K.}\ \bibnamefont {Efetov}},
  \bibinfo {author} {\bibfnamefont {P.}~\bibnamefont {Jarillo-Herrero}},
  \bibinfo {author} {\bibfnamefont {A.~H.}\ \bibnamefont {MacDonald}}, \bibinfo
  {author} {\bibfnamefont {K.~F.}\ \bibnamefont {Mak}}, \bibinfo {author}
  {\bibfnamefont {T.}~\bibnamefont {Senthil}}, \bibinfo {author} {\bibfnamefont
  {E.}~\bibnamefont {Tutuc}}, \bibinfo {author} {\bibfnamefont
  {A.}~\bibnamefont {Yazdani}}, \ and\ \bibinfo {author} {\bibfnamefont
  {A.~F.}\ \bibnamefont {Young}},\ }\bibfield  {title} {\enquote {\bibinfo
  {title} {{The marvels of moir{\'{e}} materials}},}\ }\href {\doibase
  10.1038/s41578-021-00284-1} {\bibfield  {journal} {\bibinfo  {journal} {Nat.
  Rev. Mater.}\ }\textbf {\bibinfo {volume} {6}},\ \bibinfo {pages} {201--206}
  (\bibinfo {year} {2021})}\BibitemShut {NoStop}%
\bibitem [{\citenamefont {Du}\ \emph {et~al.}(2023)\citenamefont {Du},
  \citenamefont {Molas}, \citenamefont {Huang}, \citenamefont {Zhang},
  \citenamefont {Wang},\ and\ \citenamefont {Sun}}]{Du2023}%
  \BibitemOpen
  \bibfield  {author} {\bibinfo {author} {\bibfnamefont {L.}~\bibnamefont
  {Du}}, \bibinfo {author} {\bibfnamefont {M.~R.}\ \bibnamefont {Molas}},
  \bibinfo {author} {\bibfnamefont {Z.}~\bibnamefont {Huang}}, \bibinfo
  {author} {\bibfnamefont {G.}~\bibnamefont {Zhang}}, \bibinfo {author}
  {\bibfnamefont {F.}~\bibnamefont {Wang}}, \ and\ \bibinfo {author}
  {\bibfnamefont {Z.}~\bibnamefont {Sun}},\ }\bibfield  {title} {\enquote
  {\bibinfo {title} {{Moir{\'{e}} photonics and optoelectronics}},}\ }\href
  {\doibase 10.1126/science.adg0014} {\bibfield  {journal} {\bibinfo  {journal}
  {Science}\ }\textbf {\bibinfo {volume} {379}},\ \bibinfo {pages} {eadg0014}
  (\bibinfo {year} {2023})}\BibitemShut {NoStop}%
\bibitem [{\citenamefont {Gratias}\ and\ \citenamefont
  {Quiquandon}(2023)}]{Gratias2023}%
  \BibitemOpen
  \bibfield  {author} {\bibinfo {author} {\bibfnamefont {D.}~\bibnamefont
  {Gratias}}\ and\ \bibinfo {author} {\bibfnamefont {M.}~\bibnamefont
  {Quiquandon}},\ }\bibfield  {title} {\enquote {\bibinfo {title}
  {{Crystallography of homophase twisted bilayers: coincidence, union lattices
  and space groups}},}\ }\href {\doibase 10.1107/S2053273323003662} {\bibfield
  {journal} {\bibinfo  {journal} {Acta Cryst.}\ }\textbf {\bibinfo {volume}
  {A79}},\ \bibinfo {pages} {301--317} (\bibinfo {year} {2023})}\BibitemShut
  {NoStop}%
\bibitem [{\citenamefont {Yankowitz}\ \emph {et~al.}(2019)\citenamefont
  {Yankowitz}, \citenamefont {Chen}, \citenamefont {Polshyn}, \citenamefont
  {Zhang}, \citenamefont {Watanabe}, \citenamefont {Taniguchi}, \citenamefont
  {Graf}, \citenamefont {Young},\ and\ \citenamefont {Dean}}]{Yankowitz2019}%
  \BibitemOpen
  \bibfield  {author} {\bibinfo {author} {\bibfnamefont {M.}~\bibnamefont
  {Yankowitz}}, \bibinfo {author} {\bibfnamefont {S.}~\bibnamefont {Chen}},
  \bibinfo {author} {\bibfnamefont {H.}~\bibnamefont {Polshyn}}, \bibinfo
  {author} {\bibfnamefont {Y.}~\bibnamefont {Zhang}}, \bibinfo {author}
  {\bibfnamefont {K.}~\bibnamefont {Watanabe}}, \bibinfo {author}
  {\bibfnamefont {T.}~\bibnamefont {Taniguchi}}, \bibinfo {author}
  {\bibfnamefont {D.}~\bibnamefont {Graf}}, \bibinfo {author} {\bibfnamefont
  {A.~F.}\ \bibnamefont {Young}}, \ and\ \bibinfo {author} {\bibfnamefont
  {C.~R.}\ \bibnamefont {Dean}},\ }\bibfield  {title} {\enquote {\bibinfo
  {title} {{Tuning superconductivity in twisted bilayer graphene}},}\ }\href
  {\doibase 10.1126/science.aav1910} {\bibfield  {journal} {\bibinfo  {journal}
  {Science}\ }\textbf {\bibinfo {volume} {363}},\ \bibinfo {pages} {1059--1064}
  (\bibinfo {year} {2019})}\BibitemShut {NoStop}%
\bibitem [{\citenamefont {Balents}\ \emph {et~al.}(2020)\citenamefont
  {Balents}, \citenamefont {Dean}, \citenamefont {Efetov},\ and\ \citenamefont
  {Young}}]{Balents2020}%
  \BibitemOpen
  \bibfield  {author} {\bibinfo {author} {\bibfnamefont {L.}~\bibnamefont
  {Balents}}, \bibinfo {author} {\bibfnamefont {C.~R.}\ \bibnamefont {Dean}},
  \bibinfo {author} {\bibfnamefont {D.~K.}\ \bibnamefont {Efetov}}, \ and\
  \bibinfo {author} {\bibfnamefont {A.~F.}\ \bibnamefont {Young}},\ }\bibfield
  {title} {\enquote {\bibinfo {title} {{Superconductivity and strong
  correlations in moir{\'{e}} flat bands}},}\ }\href {\doibase
  10.1038/s41567-020-0906-9} {\bibfield  {journal} {\bibinfo  {journal} {Nat.
  Phys.}\ }\textbf {\bibinfo {volume} {16}},\ \bibinfo {pages} {725--733}
  (\bibinfo {year} {2020})}\BibitemShut {NoStop}%
\bibitem [{\citenamefont {Cao}\ \emph {et~al.}(2018)\citenamefont {Cao},
  \citenamefont {Fatemi}, \citenamefont {Demir}, \citenamefont {Fang},
  \citenamefont {Tomarken}, \citenamefont {Luo}, \citenamefont
  {Sanchez-Yamagishi}, \citenamefont {Watanabe}, \citenamefont {Taniguchi},
  \citenamefont {Kaxiras}, \citenamefont {Ashoori},\ and\ \citenamefont
  {Jarillo-Herrero}}]{Cao2018}%
  \BibitemOpen
  \bibfield  {author} {\bibinfo {author} {\bibfnamefont {Y.}~\bibnamefont
  {Cao}}, \bibinfo {author} {\bibfnamefont {V.}~\bibnamefont {Fatemi}},
  \bibinfo {author} {\bibfnamefont {A.}~\bibnamefont {Demir}}, \bibinfo
  {author} {\bibfnamefont {S.}~\bibnamefont {Fang}}, \bibinfo {author}
  {\bibfnamefont {S.~L.}\ \bibnamefont {Tomarken}}, \bibinfo {author}
  {\bibfnamefont {Jason~Y.}\ \bibnamefont {Luo}}, \bibinfo {author}
  {\bibfnamefont {J.~D.}\ \bibnamefont {Sanchez-Yamagishi}}, \bibinfo {author}
  {\bibfnamefont {K.}~\bibnamefont {Watanabe}}, \bibinfo {author}
  {\bibfnamefont {T.}~\bibnamefont {Taniguchi}}, \bibinfo {author}
  {\bibfnamefont {E.}~\bibnamefont {Kaxiras}}, \bibinfo {author} {\bibfnamefont
  {R.~C.}\ \bibnamefont {Ashoori}}, \ and\ \bibinfo {author} {\bibfnamefont
  {P.}~\bibnamefont {Jarillo-Herrero}},\ }\bibfield  {title} {\enquote
  {\bibinfo {title} {{Correlated insulator behaviour at half-filling in
  magic-angle graphene superlattices}},}\ }\href {\doibase 10.1038/nature26154}
  {\bibfield  {journal} {\bibinfo  {journal} {Nature}\ }\textbf {\bibinfo
  {volume} {556}},\ \bibinfo {pages} {80--84} (\bibinfo {year}
  {2018})}\BibitemShut {NoStop}%
\bibitem [{\citenamefont {Bistritzer}\ and\ \citenamefont
  {MacDonald}(2011)}]{Bistritzer2011}%
  \BibitemOpen
  \bibfield  {author} {\bibinfo {author} {\bibfnamefont {R.}~\bibnamefont
  {Bistritzer}}\ and\ \bibinfo {author} {\bibfnamefont {A.~H.}\ \bibnamefont
  {MacDonald}},\ }\bibfield  {title} {\enquote {\bibinfo {title} {Moir\'e bands
  in twisted double-layer graphene},}\ }\href {\doibase
  10.1073/pnas.1108174108} {\bibfield  {journal} {\bibinfo  {journal} {Proc.
  Natl. Acad. Sci. USA}\ }\textbf {\bibinfo {volume} {108}},\ \bibinfo {pages}
  {12233--12237} (\bibinfo {year} {2011})}\BibitemShut {NoStop}%
\bibitem [{\citenamefont {Xian}\ \emph {et~al.}(2019)\citenamefont {Xian},
  \citenamefont {Kennes}, \citenamefont {Tancogne-Dejean}, \citenamefont
  {Altarelli},\ and\ \citenamefont {Rubio}}]{Xian2019}%
  \BibitemOpen
  \bibfield  {author} {\bibinfo {author} {\bibfnamefont {L.}~\bibnamefont
  {Xian}}, \bibinfo {author} {\bibfnamefont {D.~M.}\ \bibnamefont {Kennes}},
  \bibinfo {author} {\bibfnamefont {N.}~\bibnamefont {Tancogne-Dejean}},
  \bibinfo {author} {\bibfnamefont {M.}~\bibnamefont {Altarelli}}, \ and\
  \bibinfo {author} {\bibfnamefont {A.}~\bibnamefont {Rubio}},\ }\bibfield
  {title} {\enquote {\bibinfo {title} {{Multiflat Bands and Strong Correlations
  in Twisted Bilayer Boron Nitride: Doping-Induced Correlated Insulator and
  Superconductor}},}\ }\href {\doibase 10.1021/acs.nanolett.9b00986} {\bibfield
   {journal} {\bibinfo  {journal} {Nano Lett.}\ }\textbf {\bibinfo {volume}
  {19}},\ \bibinfo {pages} {4934} (\bibinfo {year} {2019})}\BibitemShut
  {NoStop}%
\bibitem [{\citenamefont {Kim}\ \emph {et~al.}(2023)\citenamefont {Kim},
  \citenamefont {Dominguez}, \citenamefont {Mayorga-luna}, \citenamefont {Ye},
  \citenamefont {Embley}, \citenamefont {Tan}, \citenamefont {Ni},
  \citenamefont {Liu}, \citenamefont {Ford}, \citenamefont {Gao}, \citenamefont
  {Arash}, \citenamefont {Watanabe}, \citenamefont {Taniguchi}, \citenamefont
  {Kim}, \citenamefont {Shih}, \citenamefont {Lai}, \citenamefont {Yao},
  \citenamefont {Yang}, \citenamefont {Li},\ and\ \citenamefont
  {Miyahara}}]{Kim2023}%
  \BibitemOpen
  \bibfield  {author} {\bibinfo {author} {\bibfnamefont {D.~S.}\ \bibnamefont
  {Kim}}, \bibinfo {author} {\bibfnamefont {R.~C.}\ \bibnamefont {Dominguez}},
  \bibinfo {author} {\bibfnamefont {R.}~\bibnamefont {Mayorga-luna}}, \bibinfo
  {author} {\bibfnamefont {D.}~\bibnamefont {Ye}}, \bibinfo {author}
  {\bibfnamefont {J.}~\bibnamefont {Embley}}, \bibinfo {author} {\bibfnamefont
  {T.}~\bibnamefont {Tan}}, \bibinfo {author} {\bibfnamefont {Y.}~\bibnamefont
  {Ni}}, \bibinfo {author} {\bibfnamefont {Z.}~\bibnamefont {Liu}}, \bibinfo
  {author} {\bibfnamefont {M.}~\bibnamefont {Ford}}, \bibinfo {author}
  {\bibfnamefont {F.~Y.}\ \bibnamefont {Gao}}, \bibinfo {author} {\bibfnamefont
  {S.}~\bibnamefont {Arash}}, \bibinfo {author} {\bibfnamefont
  {K.}~\bibnamefont {Watanabe}}, \bibinfo {author} {\bibfnamefont
  {T.}~\bibnamefont {Taniguchi}}, \bibinfo {author} {\bibfnamefont
  {S.}~\bibnamefont {Kim}}, \bibinfo {author} {\bibfnamefont {C.-K.}\
  \bibnamefont {Shih}}, \bibinfo {author} {\bibfnamefont {K.}~\bibnamefont
  {Lai}}, \bibinfo {author} {\bibfnamefont {W.}~\bibnamefont {Yao}}, \bibinfo
  {author} {\bibfnamefont {L.}~\bibnamefont {Yang}}, \bibinfo {author}
  {\bibfnamefont {X.}~\bibnamefont {Li}}, \ and\ \bibinfo {author}
  {\bibfnamefont {Y.}~\bibnamefont {Miyahara}},\ }\bibfield  {title} {\enquote
  {\bibinfo {title} {{Electrostatic moir{\'{e}} potential from twisted
  hexagonal boron nitride layers}},}\ }\href {\doibase
  10.1038/s41563-023-01637-7} {\bibfield  {journal} {\bibinfo  {journal} {Nat
  Mater.}\ } (\bibinfo {year} {2023}),\ 10.1038/s41563-023-01637-7}\BibitemShut
  {NoStop}%
\bibitem [{\citenamefont {Scuri}\ \emph {et~al.}(2020)\citenamefont {Scuri},
  \citenamefont {Andersen}, \citenamefont {Zhou}, \citenamefont {Wild},
  \citenamefont {Sung}, \citenamefont {Gelly}, \citenamefont {B\'erub\'e},
  \citenamefont {Heo}, \citenamefont {Shao}, \citenamefont {Joe}, \citenamefont
  {Mier~Valdivia}, \citenamefont {Taniguchi}, \citenamefont {Watanabe},
  \citenamefont {Lon\u{c}ar}, \citenamefont {Kim}, \citenamefont {Lukin},\ and\
  \citenamefont {Park}}]{Scuri2020}%
  \BibitemOpen
  \bibfield  {author} {\bibinfo {author} {\bibfnamefont {G.}~\bibnamefont
  {Scuri}}, \bibinfo {author} {\bibfnamefont {T.~I.}\ \bibnamefont {Andersen}},
  \bibinfo {author} {\bibfnamefont {Y.}~\bibnamefont {Zhou}}, \bibinfo {author}
  {\bibfnamefont {D.~S.}\ \bibnamefont {Wild}}, \bibinfo {author}
  {\bibfnamefont {J.}~\bibnamefont {Sung}}, \bibinfo {author} {\bibfnamefont
  {R.~J.}\ \bibnamefont {Gelly}}, \bibinfo {author} {\bibfnamefont
  {D.}~\bibnamefont {B\'erub\'e}}, \bibinfo {author} {\bibfnamefont
  {H.}~\bibnamefont {Heo}}, \bibinfo {author} {\bibfnamefont {L.}~\bibnamefont
  {Shao}}, \bibinfo {author} {\bibfnamefont {A.~Y.}\ \bibnamefont {Joe}},
  \bibinfo {author} {\bibfnamefont {A.~M.}\ \bibnamefont {Mier~Valdivia}},
  \bibinfo {author} {\bibfnamefont {T.}~\bibnamefont {Taniguchi}}, \bibinfo
  {author} {\bibfnamefont {K.}~\bibnamefont {Watanabe}}, \bibinfo {author}
  {\bibfnamefont {M.}~\bibnamefont {Lon\u{c}ar}}, \bibinfo {author}
  {\bibfnamefont {P.}~\bibnamefont {Kim}}, \bibinfo {author} {\bibfnamefont
  {M.~D.}\ \bibnamefont {Lukin}}, \ and\ \bibinfo {author} {\bibfnamefont
  {H.}~\bibnamefont {Park}},\ }\bibfield  {title} {\enquote {\bibinfo {title}
  {Electrically tunable valley dynamics in twisted {WS}e$_{2}$/{WS}e$_{2}$
  bilayers},}\ }\href {\doibase 10.1103/PhysRevLett.124.217403} {\bibfield
  {journal} {\bibinfo  {journal} {Phys. Rev. Lett.}\ }\textbf {\bibinfo
  {volume} {124}},\ \bibinfo {pages} {217403} (\bibinfo {year}
  {2020})}\BibitemShut {NoStop}%
\bibitem [{\citenamefont {Zhao}\ \emph {et~al.}(2020)\citenamefont {Zhao},
  \citenamefont {Yang}, \citenamefont {Zhang},\ and\ \citenamefont
  {Wei}}]{Zhao2020}%
  \BibitemOpen
  \bibfield  {author} {\bibinfo {author} {\bibfnamefont {X.-J.}\ \bibnamefont
  {Zhao}}, \bibinfo {author} {\bibfnamefont {Y.}~\bibnamefont {Yang}}, \bibinfo
  {author} {\bibfnamefont {D.-B.}\ \bibnamefont {Zhang}}, \ and\ \bibinfo
  {author} {\bibfnamefont {S.-H.}\ \bibnamefont {Wei}},\ }\bibfield  {title}
  {\enquote {\bibinfo {title} {Formation of bloch flat bands in polar twisted
  bilayers without magic angles},}\ }\href {\doibase
  10.1103/PhysRevLett.124.086401} {\bibfield  {journal} {\bibinfo  {journal}
  {Phys. Rev. Lett.}\ }\textbf {\bibinfo {volume} {124}},\ \bibinfo {pages}
  {086401} (\bibinfo {year} {2020})}\BibitemShut {NoStop}%
\bibitem [{\citenamefont {Walet}\ and\ \citenamefont
  {Guinea}(2021)}]{Walet2021}%
  \BibitemOpen
  \bibfield  {author} {\bibinfo {author} {\bibfnamefont {N.~R.}\ \bibnamefont
  {Walet}}\ and\ \bibinfo {author} {\bibfnamefont {F.}~\bibnamefont {Guinea}},\
  }\bibfield  {title} {\enquote {\bibinfo {title} {{Flat bands, strains, and
  charge distribution in twisted bilayer h-BN}},}\ }\href {\doibase
  10.1103/PhysRevB.103.125427} {\bibfield  {journal} {\bibinfo  {journal}
  {Phys. Rev. B}\ }\textbf {\bibinfo {volume} {103}},\ \bibinfo {pages}
  {125427} (\bibinfo {year} {2021})}\BibitemShut {NoStop}%
\bibitem [{\citenamefont {Latil}\ \emph {et~al.}(2023)\citenamefont {Latil},
  \citenamefont {Amara},\ and\ \citenamefont {Sponza}}]{Latil2023}%
  \BibitemOpen
  \bibfield  {author} {\bibinfo {author} {\bibfnamefont {S.}~\bibnamefont
  {Latil}}, \bibinfo {author} {\bibfnamefont {H.}~\bibnamefont {Amara}}, \ and\
  \bibinfo {author} {\bibfnamefont {L.}~\bibnamefont {Sponza}},\ }\bibfield
  {title} {\enquote {\bibinfo {title} {{Structural classification of boron
  nitride twisted bilayers and ab initio investigation of their
  stacking-dependent electronic structure}},}\ }\href {\doibase
  10.21468/SciPostPhys.14.3.053} {\bibfield  {journal} {\bibinfo  {journal}
  {SciPost Phys.}\ }\textbf {\bibinfo {volume} {14}},\ \bibinfo {pages} {053}
  (\bibinfo {year} {2023})}\BibitemShut {NoStop}%
\bibitem [{\citenamefont {Scheer}\ \emph {et~al.}(2022)\citenamefont {Scheer},
  \citenamefont {Gu},\ and\ \citenamefont {Lian}}]{Scheer2022}%
  \BibitemOpen
  \bibfield  {author} {\bibinfo {author} {\bibfnamefont {M.~G.}\ \bibnamefont
  {Scheer}}, \bibinfo {author} {\bibfnamefont {K.}~\bibnamefont {Gu}}, \ and\
  \bibinfo {author} {\bibfnamefont {B.}~\bibnamefont {Lian}},\ }\bibfield
  {title} {\enquote {\bibinfo {title} {Magic angles in twisted bilayer graphene
  near commensuration: Towards a hypermagic regime},}\ }\href {\doibase
  10.1103/PhysRevB.106.115418} {\bibfield  {journal} {\bibinfo  {journal}
  {Phys. Rev. B}\ }\textbf {\bibinfo {volume} {106}},\ \bibinfo {pages}
  {115418} (\bibinfo {year} {2022})}\BibitemShut {NoStop}%
\bibitem [{\citenamefont {Fang}\ \emph {et~al.}(2015)\citenamefont {Fang},
  \citenamefont {Defo}, \citenamefont {Shirodkar}, \citenamefont {Lieu},
  \citenamefont {Tritsaris},\ and\ \citenamefont {Kaxiras}}]{Fang2015}%
  \BibitemOpen
  \bibfield  {author} {\bibinfo {author} {\bibfnamefont {S.}~\bibnamefont
  {Fang}}, \bibinfo {author} {\bibfnamefont {R.~K.}\ \bibnamefont {Defo}},
  \bibinfo {author} {\bibfnamefont {S.~N.}\ \bibnamefont {Shirodkar}}, \bibinfo
  {author} {\bibfnamefont {S.}~\bibnamefont {Lieu}}, \bibinfo {author}
  {\bibfnamefont {G.~A.}\ \bibnamefont {Tritsaris}}, \ and\ \bibinfo {author}
  {\bibfnamefont {E.}~\bibnamefont {Kaxiras}},\ }\bibfield  {title} {\enquote
  {\bibinfo {title} {Ab initio tight-binding hamiltonian for transition metal
  dichalcogenides},}\ }\href {\doibase 10.1103/PhysRevB.92.205108} {\bibfield
  {journal} {\bibinfo  {journal} {Phys. Rev. B}\ }\textbf {\bibinfo {volume}
  {92}},\ \bibinfo {pages} {205108} (\bibinfo {year} {2015})}\BibitemShut
  {NoStop}%
\bibitem [{\citenamefont {Trambly~de Laissardi\`ere}\ \emph
  {et~al.}(2010)\citenamefont {Trambly~de Laissardi\`ere}, \citenamefont
  {Mayou},\ and\ \citenamefont {Magaud}}]{Trambly2010}%
  \BibitemOpen
  \bibfield  {author} {\bibinfo {author} {\bibfnamefont {G.}~\bibnamefont
  {Trambly~de Laissardi\`ere}}, \bibinfo {author} {\bibfnamefont
  {D.}~\bibnamefont {Mayou}}, \ and\ \bibinfo {author} {\bibfnamefont
  {L.}~\bibnamefont {Magaud}},\ }\bibfield  {title} {\enquote {\bibinfo {title}
  {Localization of {Dirac} electrons in rotated graphene bilayers},}\ }\href
  {\doibase 10.1021/nl902948m} {\bibfield  {journal} {\bibinfo  {journal} {Nano
  Lett.}\ }\textbf {\bibinfo {volume} {10}},\ \bibinfo {pages} {804--808}
  (\bibinfo {year} {2010})}\BibitemShut {NoStop}%
\bibitem [{\citenamefont {Venkateswarlu}\ \emph {et~al.}(2020)\citenamefont
  {Venkateswarlu}, \citenamefont {Honecker},\ and\ \citenamefont {Trambly~de
  Laissardi\`ere}}]{Venkateswarlu2020}%
  \BibitemOpen
  \bibfield  {author} {\bibinfo {author} {\bibfnamefont {S.}~\bibnamefont
  {Venkateswarlu}}, \bibinfo {author} {\bibfnamefont {A.}~\bibnamefont
  {Honecker}}, \ and\ \bibinfo {author} {\bibfnamefont {G.}~\bibnamefont
  {Trambly~de Laissardi\`ere}},\ }\bibfield  {title} {\enquote {\bibinfo
  {title} {Electronic localization in twisted bilayer {M}o{S}$_{2}$ with small
  rotation angle},}\ }\href {\doibase 10.1103/PhysRevB.102.081103} {\bibfield
  {journal} {\bibinfo  {journal} {Phys. Rev. B}\ }\textbf {\bibinfo {volume}
  {102}},\ \bibinfo {pages} {081103} (\bibinfo {year} {2020})}\BibitemShut
  {NoStop}%
\bibitem [{\citenamefont {Long}\ \emph {et~al.}(2022)\citenamefont {Long},
  \citenamefont {Pantale{\'{o}}n}, \citenamefont {Zhan}, \citenamefont
  {Guinea}, \citenamefont {Silva-Guill{\'{e}}n},\ and\ \citenamefont
  {Yuan}}]{Long2022}%
  \BibitemOpen
  \bibfield  {author} {\bibinfo {author} {\bibfnamefont {M.}~\bibnamefont
  {Long}}, \bibinfo {author} {\bibfnamefont {P.~A.}\ \bibnamefont
  {Pantale{\'{o}}n}}, \bibinfo {author} {\bibfnamefont {Z.}~\bibnamefont
  {Zhan}}, \bibinfo {author} {\bibfnamefont {F.}~\bibnamefont {Guinea}},
  \bibinfo {author} {\bibfnamefont {J.~A.}\ \bibnamefont
  {Silva-Guill{\'{e}}n}}, \ and\ \bibinfo {author} {\bibfnamefont
  {S.}~\bibnamefont {Yuan}},\ }\bibfield  {title} {\enquote {\bibinfo {title}
  {{An atomistic approach for the structural and electronic properties of
  twisted bilayer graphene-boron nitride heterostructures}},}\ }\href {\doibase
  10.1038/s41524-022-00763-1} {\bibfield  {journal} {\bibinfo  {journal} {npj
  Comput. Mater.}\ }\textbf {\bibinfo {volume} {8}},\ \bibinfo {pages} {73}
  (\bibinfo {year} {2022})}\BibitemShut {NoStop}%
\bibitem [{\citenamefont {Ochoa}\ and\ \citenamefont
  {Asenjo-Garcia}(2020)}]{Ochoa2020}%
  \BibitemOpen
  \bibfield  {author} {\bibinfo {author} {\bibfnamefont {H.}~\bibnamefont
  {Ochoa}}\ and\ \bibinfo {author} {\bibfnamefont {A.}~\bibnamefont
  {Asenjo-Garcia}},\ }\bibfield  {title} {\enquote {\bibinfo {title} {{Flat
  Bands and Chiral Optical Response of Moir{\'{e}} Insulators}},}\ }\href
  {\doibase 10.1103/PhysRevLett.125.037402} {\bibfield  {journal} {\bibinfo
  {journal} {Phys. Rev. Lett.}\ }\textbf {\bibinfo {volume} {125}},\ \bibinfo
  {pages} {37402} (\bibinfo {year} {2020})},\ \Eprint
  {http://arxiv.org/abs/2002.09804} {arXiv:2002.09804} \BibitemShut {NoStop}%
\bibitem [{\citenamefont {Sboychakov}\ \emph {et~al.}(2015)\citenamefont
  {Sboychakov}, \citenamefont {Rakhmanov}, \citenamefont {Rozhkov},\ and\
  \citenamefont {Nori}}]{Sboychakov2015}%
  \BibitemOpen
  \bibfield  {author} {\bibinfo {author} {\bibfnamefont {A.~O.}\ \bibnamefont
  {Sboychakov}}, \bibinfo {author} {\bibfnamefont {A.~L.}\ \bibnamefont
  {Rakhmanov}}, \bibinfo {author} {\bibfnamefont {A.~V.}\ \bibnamefont
  {Rozhkov}}, \ and\ \bibinfo {author} {\bibfnamefont {Franco}\ \bibnamefont
  {Nori}},\ }\bibfield  {title} {\enquote {\bibinfo {title} {Electronic
  spectrum of twisted bilayer graphene},}\ }\href {\doibase
  10.1103/PhysRevB.92.075402} {\bibfield  {journal} {\bibinfo  {journal} {Phys.
  Rev. B}\ }\textbf {\bibinfo {volume} {92}},\ \bibinfo {pages} {075402}
  (\bibinfo {year} {2015})}\BibitemShut {NoStop}%
\bibitem [{\citenamefont {Pal}\ \emph {et~al.}(2019)\citenamefont {Pal},
  \citenamefont {Spitz},\ and\ \citenamefont {Kindermann}}]{Pal2019}%
  \BibitemOpen
  \bibfield  {author} {\bibinfo {author} {\bibfnamefont {H.~K.}\ \bibnamefont
  {Pal}}, \bibinfo {author} {\bibfnamefont {S.}~\bibnamefont {Spitz}}, \ and\
  \bibinfo {author} {\bibfnamefont {M.}~\bibnamefont {Kindermann}},\ }\bibfield
   {title} {\enquote {\bibinfo {title} {Emergent geometric frustration and flat
  band in moir\'e bilayer graphene},}\ }\href {\doibase
  10.1103/PhysRevLett.123.186402} {\bibfield  {journal} {\bibinfo  {journal}
  {Phys. Rev. Lett.}\ }\textbf {\bibinfo {volume} {123}},\ \bibinfo {pages}
  {186402} (\bibinfo {year} {2019})}\BibitemShut {NoStop}%
\bibitem [{\citenamefont {Mondal}\ \emph {et~al.}(2023)\citenamefont {Mondal},
  \citenamefont {Ghadimi},\ and\ \citenamefont {Yang}}]{Mondal2023}%
  \BibitemOpen
  \bibfield  {author} {\bibinfo {author} {\bibfnamefont {C.}~\bibnamefont
  {Mondal}}, \bibinfo {author} {\bibfnamefont {R.}~\bibnamefont {Ghadimi}}, \
  and\ \bibinfo {author} {\bibfnamefont {B.-J.}\ \bibnamefont {Yang}},\
  }\bibfield  {title} {\enquote {\bibinfo {title} {Quantum valley and subvalley
  hall effect in large-angle twisted bilayer graphene},}\ }\href {\doibase
  10.1103/PhysRevB.108.L121405} {\bibfield  {journal} {\bibinfo  {journal}
  {Phys. Rev. B}\ }\textbf {\bibinfo {volume} {108}},\ \bibinfo {pages}
  {L121405} (\bibinfo {year} {2023})}\BibitemShut {NoStop}%
\bibitem [{\citenamefont {Ahn}\ \emph {et~al.}(2018)\citenamefont {Ahn},
  \citenamefont {Moon}, \citenamefont {Kim}, \citenamefont {Kim}, \citenamefont
  {Shin}, \citenamefont {Kim}, \citenamefont {Cha}, \citenamefont {Kahng},
  \citenamefont {Kim}, \citenamefont {Koshino}, \citenamefont {Son},
  \citenamefont {Yang},\ and\ \citenamefont {Ahn}}]{Ahn2018}%
  \BibitemOpen
  \bibfield  {author} {\bibinfo {author} {\bibfnamefont {S.~J.}\ \bibnamefont
  {Ahn}}, \bibinfo {author} {\bibfnamefont {P.}~\bibnamefont {Moon}}, \bibinfo
  {author} {\bibfnamefont {T.-H.}\ \bibnamefont {Kim}}, \bibinfo {author}
  {\bibfnamefont {H.-W.}\ \bibnamefont {Kim}}, \bibinfo {author} {\bibfnamefont
  {H.-C.}\ \bibnamefont {Shin}}, \bibinfo {author} {\bibfnamefont {E.~H.}\
  \bibnamefont {Kim}}, \bibinfo {author} {\bibfnamefont {H.~W.}\ \bibnamefont
  {Cha}}, \bibinfo {author} {\bibfnamefont {S.-J.}\ \bibnamefont {Kahng}},
  \bibinfo {author} {\bibfnamefont {P.}~\bibnamefont {Kim}}, \bibinfo {author}
  {\bibfnamefont {M.}~\bibnamefont {Koshino}}, \bibinfo {author} {\bibfnamefont
  {Y.-W.}\ \bibnamefont {Son}}, \bibinfo {author} {\bibfnamefont {C.-W.}\
  \bibnamefont {Yang}}, \ and\ \bibinfo {author} {\bibfnamefont {J.~R.}\
  \bibnamefont {Ahn}},\ }\bibfield  {title} {\enquote {\bibinfo {title} {Dirac
  electrons in a dodecagonal graphene quasicrystal},}\ }\href {\doibase
  10.1126/science.aar8412} {\bibfield  {journal} {\bibinfo  {journal}
  {Science}\ }\textbf {\bibinfo {volume} {361}},\ \bibinfo {pages} {782--786}
  (\bibinfo {year} {2018})}\BibitemShut {NoStop}%
\bibitem [{\citenamefont {Yao}\ \emph {et~al.}(2018)\citenamefont {Yao},
  \citenamefont {Wang}, \citenamefont {Bao}, \citenamefont {Zhang},
  \citenamefont {Zhang}, \citenamefont {Bao}, \citenamefont {Chan},
  \citenamefont {Chen}, \citenamefont {Avila}, \citenamefont {Asensio},
  \citenamefont {Zhu},\ and\ \citenamefont {Zhou}}]{Yao2018}%
  \BibitemOpen
  \bibfield  {author} {\bibinfo {author} {\bibfnamefont {W.}~\bibnamefont
  {Yao}}, \bibinfo {author} {\bibfnamefont {E.}~\bibnamefont {Wang}}, \bibinfo
  {author} {\bibfnamefont {C.}~\bibnamefont {Bao}}, \bibinfo {author}
  {\bibfnamefont {Y.}~\bibnamefont {Zhang}}, \bibinfo {author} {\bibfnamefont
  {K.}~\bibnamefont {Zhang}}, \bibinfo {author} {\bibfnamefont
  {K.}~\bibnamefont {Bao}}, \bibinfo {author} {\bibfnamefont {C.~K.}\
  \bibnamefont {Chan}}, \bibinfo {author} {\bibfnamefont {C.}~\bibnamefont
  {Chen}}, \bibinfo {author} {\bibfnamefont {J.}~\bibnamefont {Avila}},
  \bibinfo {author} {\bibfnamefont {M.~C.}\ \bibnamefont {Asensio}}, \bibinfo
  {author} {\bibfnamefont {J.}~\bibnamefont {Zhu}}, \ and\ \bibinfo {author}
  {\bibfnamefont {S.}~\bibnamefont {Zhou}},\ }\bibfield  {title} {\enquote
  {\bibinfo {title} {Quasicrystalline 30° twisted bilayer graphene as an
  incommensurate superlattice with strong interlayer coupling},}\ }\href
  {\doibase 10.1073/pnas.1720865115} {\bibfield  {journal} {\bibinfo  {journal}
  {Proc. Natl. Acad. Sci. USA}\ }\textbf {\bibinfo {volume} {115}},\ \bibinfo
  {pages} {6928--6933} (\bibinfo {year} {2018})}\BibitemShut {NoStop}%
\bibitem [{\citenamefont {Moon}\ \emph {et~al.}(2019)\citenamefont {Moon},
  \citenamefont {Koshino},\ and\ \citenamefont {Son}}]{Moon2019}%
  \BibitemOpen
  \bibfield  {author} {\bibinfo {author} {\bibfnamefont {P.}~\bibnamefont
  {Moon}}, \bibinfo {author} {\bibfnamefont {M.}~\bibnamefont {Koshino}}, \
  and\ \bibinfo {author} {\bibfnamefont {Y.-W.}\ \bibnamefont {Son}},\
  }\bibfield  {title} {\enquote {\bibinfo {title} {Quasicrystalline electronic
  states in ${30}^{\ensuremath{\circ}}$ rotated twisted bilayer graphene},}\
  }\href {\doibase 10.1103/PhysRevB.99.165430} {\bibfield  {journal} {\bibinfo
  {journal} {Phys. Rev. B}\ }\textbf {\bibinfo {volume} {99}},\ \bibinfo
  {pages} {165430} (\bibinfo {year} {2019})}\BibitemShut {NoStop}%
\bibitem [{\citenamefont {Chernozatonskii}\ and\ \citenamefont
  {Kochaev}(2023)}]{Chernozatonskii2023}%
  \BibitemOpen
  \bibfield  {author} {\bibinfo {author} {\bibfnamefont {L.~A.}\ \bibnamefont
  {Chernozatonskii}}\ and\ \bibinfo {author} {\bibfnamefont {A.~I.}\
  \bibnamefont {Kochaev}},\ }\bibfield  {title} {\enquote {\bibinfo {title}
  {{BN Diamane-like Quasicrystal Based on 30° Twisted h-BN Bilayers and Its
  Approximants: Features of the Atomic Structure and Electronic Properties}},}\
  }\href {\doibase 10.3390/cryst13030421} {\bibfield  {journal} {\bibinfo
  {journal} {Crystals}\ }\textbf {\bibinfo {volume} {13}},\ \bibinfo {pages}
  {421} (\bibinfo {year} {2023})}\BibitemShut {NoStop}%
\bibitem [{Sup()}]{SupMat}%
  \BibitemOpen
  \href@noop {} {}\bibinfo {note} {See Supplemental Material [url] for details
  on the nomenclature defining the structures, on the DFT calculation
  parameters, on the tight-binding models and for complementary results and
  analysis.}\BibitemShut {Stop}%
\bibitem [{\citenamefont {Trambly~de Laissardi\`ere}\ \emph
  {et~al.}(2012)\citenamefont {Trambly~de Laissardi\`ere}, \citenamefont
  {Mayou},\ and\ \citenamefont {Magaud}}]{Trambly2012}%
  \BibitemOpen
  \bibfield  {author} {\bibinfo {author} {\bibfnamefont {G.}~\bibnamefont
  {Trambly~de Laissardi\`ere}}, \bibinfo {author} {\bibfnamefont
  {D.}~\bibnamefont {Mayou}}, \ and\ \bibinfo {author} {\bibfnamefont
  {L.}~\bibnamefont {Magaud}},\ }\bibfield  {title} {\enquote {\bibinfo {title}
  {Numerical studies of confined states in rotated bilayers of graphene},}\
  }\href {\doibase 10.1103/PhysRevB.86.125413} {\bibfield  {journal} {\bibinfo
  {journal} {Phys. Rev. B}\ }\textbf {\bibinfo {volume} {86}},\ \bibinfo
  {pages} {125413} (\bibinfo {year} {2012})}\BibitemShut {NoStop}%
\bibitem [{\citenamefont {Henriques}\ \emph {et~al.}(2022)\citenamefont
  {Henriques}, \citenamefont {Amorim}, \citenamefont {Ribeiro},\ and\
  \citenamefont {Peres}}]{Henriques2022}%
  \BibitemOpen
  \bibfield  {author} {\bibinfo {author} {\bibfnamefont {J.~C.~G.}\
  \bibnamefont {Henriques}}, \bibinfo {author} {\bibfnamefont {B.}~\bibnamefont
  {Amorim}}, \bibinfo {author} {\bibfnamefont {R.~M.}\ \bibnamefont {Ribeiro}},
  \ and\ \bibinfo {author} {\bibfnamefont {N.~M.~R.}\ \bibnamefont {Peres}},\
  }\bibfield  {title} {\enquote {\bibinfo {title} {Excitonic response of {AA}'
  and {AB} stacked hbn bilayers},}\ }\href {\doibase
  10.1103/PhysRevB.105.115421} {\bibfield  {journal} {\bibinfo  {journal}
  {Phys. Rev. B}\ }\textbf {\bibinfo {volume} {105}},\ \bibinfo {pages}
  {115421} (\bibinfo {year} {2022})}\BibitemShut {NoStop}%
\bibitem [{\citenamefont {Roman-Taboada}\ \emph {et~al.}(2023)\citenamefont
  {Roman-Taboada}, \citenamefont {Obregon-Castillo}, \citenamefont
  {Botello-Mendez},\ and\ \citenamefont {Noguez}}]{Roman-Taboada2023}%
  \BibitemOpen
  \bibfield  {author} {\bibinfo {author} {\bibfnamefont {P.}~\bibnamefont
  {Roman-Taboada}}, \bibinfo {author} {\bibfnamefont {E.}~\bibnamefont
  {Obregon-Castillo}}, \bibinfo {author} {\bibfnamefont {A.~R.}\ \bibnamefont
  {Botello-Mendez}}, \ and\ \bibinfo {author} {\bibfnamefont {C.}~\bibnamefont
  {Noguez}},\ }\bibfield  {title} {\enquote {\bibinfo {title} {Excitons in
  twisted {AA}' hexagonal boron nitride bilayers},}\ }\href {\doibase
  10.1103/PhysRevB.108.075109} {\bibfield  {journal} {\bibinfo  {journal}
  {Phys. Rev. B}\ }\textbf {\bibinfo {volume} {108}},\ \bibinfo {pages}
  {075109} (\bibinfo {year} {2023})}\BibitemShut {NoStop}%
\bibitem [{\citenamefont {Fujiwara}(1990)}]{Fujiwara90}%
  \BibitemOpen
  \bibfield  {author} {\bibinfo {author} {\bibfnamefont {T.}~\bibnamefont
  {Fujiwara}},\ }\bibfield  {title} {\enquote {\bibinfo {title} {Electronic
  structure in three-dimensional quasicrystals},}\ }\href {\doibase
  https://doi.org/10.1016/0022-3093(90)90660-E} {\bibfield  {journal} {\bibinfo
   {journal} {J. Non-Cryst. Solids}\ }\textbf {\bibinfo {volume} {117-118}},\
  \bibinfo {pages} {844--847} (\bibinfo {year} {1990})}\BibitemShut {NoStop}%
\bibitem [{\citenamefont {Roche}\ \emph {et~al.}(1997)\citenamefont {Roche},
  \citenamefont {Trambly~de Laissardière},\ and\ \citenamefont
  {Mayou}}]{Roche97}%
  \BibitemOpen
  \bibfield  {author} {\bibinfo {author} {\bibfnamefont {S.}~\bibnamefont
  {Roche}}, \bibinfo {author} {\bibfnamefont {G.}~\bibnamefont {Trambly~de
  Laissardière}}, \ and\ \bibinfo {author} {\bibfnamefont {D.}~\bibnamefont
  {Mayou}},\ }\bibfield  {title} {\enquote {\bibinfo {title} {{Electronic
  transport properties of quasicrystals}},}\ }\href {\doibase 10.1063/1.531914}
  {\bibfield  {journal} {\bibinfo  {journal} {J. Math. Phys.}\ }\textbf
  {\bibinfo {volume} {38}},\ \bibinfo {pages} {1794--1822} (\bibinfo {year}
  {1997})},\ \Eprint
  {http://arxiv.org/abs/https://pubs.aip.org/aip/jmp/article-pdf/38/4/1794/19255683/1794\_1\_online.pdf}
  {https://pubs.aip.org/aip/jmp/article-pdf/38/4/1794/19255683/1794\_1\_online.pdf}
  \BibitemShut {NoStop}%
\bibitem [{\citenamefont {Krajcí}\ and\ \citenamefont
  {Hafner}(2001)}]{Krajci2001}%
  \BibitemOpen
  \bibfield  {author} {\bibinfo {author} {\bibfnamefont {M.}~\bibnamefont
  {Krajcí}}\ and\ \bibinfo {author} {\bibfnamefont {J.}~\bibnamefont
  {Hafner}},\ }\bibfield  {title} {\enquote {\bibinfo {title} {Fermi surfaces
  and electronic transport properties of quasicrystalline approximants},}\
  }\href {\doibase 10.1088/0953-8984/13/17/302} {\bibfield  {journal} {\bibinfo
   {journal} {Journal of Physics: Condensed Matter}\ }\textbf {\bibinfo
  {volume} {13}},\ \bibinfo {pages} {3817} (\bibinfo {year}
  {2001})}\BibitemShut {NoStop}%
\bibitem [{\citenamefont {Yu}\ \emph {et~al.}(2019)\citenamefont {Yu},
  \citenamefont {Wu}, \citenamefont {Zhan}, \citenamefont {Katsnelson},\ and\
  \citenamefont {Yuan}}]{Yu2019}%
  \BibitemOpen
  \bibfield  {author} {\bibinfo {author} {\bibfnamefont {Guodong}\ \bibnamefont
  {Yu}}, \bibinfo {author} {\bibfnamefont {Zewen}\ \bibnamefont {Wu}}, \bibinfo
  {author} {\bibfnamefont {Zhen}\ \bibnamefont {Zhan}}, \bibinfo {author}
  {\bibfnamefont {Mikhail~I.}\ \bibnamefont {Katsnelson}}, \ and\ \bibinfo
  {author} {\bibfnamefont {Shengjun}\ \bibnamefont {Yuan}},\ }\bibfield
  {title} {\enquote {\bibinfo {title} {{Dodecagonal bilayer graphene
  quasicrystal and its approximants}},}\ }\href {\doibase
  10.1038/s41524-019-0258-0} {\bibfield  {journal} {\bibinfo  {journal} {npj
  Computational Materials}\ }\textbf {\bibinfo {volume} {5}},\ \bibinfo {pages}
  {122} (\bibinfo {year} {2019})},\ \Eprint {http://arxiv.org/abs/1907.08792}
  {arXiv:1907.08792} \BibitemShut {NoStop}%
\bibitem [{\citenamefont {Uri}\ \emph {et~al.}(2023)\citenamefont {Uri},
  \citenamefont {de~la Barrera}, \citenamefont {Randeria}, \citenamefont
  {Rodan-Legrain}, \citenamefont {Devakul}, \citenamefont {Crowley},
  \citenamefont {Paul}, \citenamefont {Watanabe}, \citenamefont {Taniguchi},
  \citenamefont {Lifshitz}, \citenamefont {Fu}, \citenamefont {Ashoori},\ and\
  \citenamefont {Jarillo-Herrero}}]{Uri2023a}%
  \BibitemOpen
  \bibfield  {author} {\bibinfo {author} {\bibfnamefont {Aviram}\ \bibnamefont
  {Uri}}, \bibinfo {author} {\bibfnamefont {Sergio~C.}\ \bibnamefont {de~la
  Barrera}}, \bibinfo {author} {\bibfnamefont {Mallika~T.}\ \bibnamefont
  {Randeria}}, \bibinfo {author} {\bibfnamefont {Daniel}\ \bibnamefont
  {Rodan-Legrain}}, \bibinfo {author} {\bibfnamefont {Trithep}\ \bibnamefont
  {Devakul}}, \bibinfo {author} {\bibfnamefont {Philip~J.D.}\ \bibnamefont
  {Crowley}}, \bibinfo {author} {\bibfnamefont {Nisarga}\ \bibnamefont {Paul}},
  \bibinfo {author} {\bibfnamefont {Kenji}\ \bibnamefont {Watanabe}}, \bibinfo
  {author} {\bibfnamefont {Takashi}\ \bibnamefont {Taniguchi}}, \bibinfo
  {author} {\bibfnamefont {Ron}\ \bibnamefont {Lifshitz}}, \bibinfo {author}
  {\bibfnamefont {Liang}\ \bibnamefont {Fu}}, \bibinfo {author} {\bibfnamefont
  {Raymond~C.}\ \bibnamefont {Ashoori}}, \ and\ \bibinfo {author}
  {\bibfnamefont {Pablo}\ \bibnamefont {Jarillo-Herrero}},\ }\bibfield  {title}
  {\enquote {\bibinfo {title} {{Superconductivity and strong interactions in a
  tunable moir{\'{e}} quasicrystal}},}\ }\href {\doibase
  10.1038/s41586-023-06294-z} {\bibfield  {journal} {\bibinfo  {journal}
  {Nature}\ }\textbf {\bibinfo {volume} {620}},\ \bibinfo {pages} {762--767}
  (\bibinfo {year} {2023})}\BibitemShut {NoStop}%
\bibitem [{\citenamefont {Galvani}\ \emph {et~al.}(2016)\citenamefont
  {Galvani}, \citenamefont {Paleari}, \citenamefont {Miranda}, \citenamefont
  {Molina-S{\'{a}}nchez}, \citenamefont {Wirtz}, \citenamefont {Latil},
  \citenamefont {Amara},\ and\ \citenamefont {Ducastelle}}]{Galvani2016}%
  \BibitemOpen
  \bibfield  {author} {\bibinfo {author} {\bibfnamefont {T.}~\bibnamefont
  {Galvani}}, \bibinfo {author} {\bibfnamefont {F.}~\bibnamefont {Paleari}},
  \bibinfo {author} {\bibfnamefont {H.}~\bibnamefont {Miranda}}, \bibinfo
  {author} {\bibfnamefont {A.}~\bibnamefont {Molina-S{\'{a}}nchez}}, \bibinfo
  {author} {\bibfnamefont {L.}~\bibnamefont {Wirtz}}, \bibinfo {author}
  {\bibfnamefont {S.}~\bibnamefont {Latil}}, \bibinfo {author} {\bibfnamefont
  {H.}~\bibnamefont {Amara}}, \ and\ \bibinfo {author} {\bibfnamefont
  {F}~\bibnamefont {Ducastelle}},\ }\bibfield  {title} {\enquote {\bibinfo
  {title} {{Excitons in boron nitride single layer}},}\ }\href {\doibase
  10.1103/PhysRevB.94.125303} {\bibfield  {journal} {\bibinfo  {journal} {Phys.
  Rev. B}\ }\textbf {\bibinfo {volume} {94}},\ \bibinfo {pages} {125303}
  (\bibinfo {year} {2016})}\BibitemShut {NoStop}%
\bibitem [{\citenamefont {Yu}\ \emph {et~al.}(2023)\citenamefont {Yu},
  \citenamefont {Wang}, \citenamefont {Katsnelson},\ and\ \citenamefont
  {Yuan}}]{Yu2023}%
  \BibitemOpen
  \bibfield  {author} {\bibinfo {author} {\bibfnamefont {G.}~\bibnamefont
  {Yu}}, \bibinfo {author} {\bibfnamefont {Y.}~\bibnamefont {Wang}}, \bibinfo
  {author} {\bibfnamefont {M.~I.}\ \bibnamefont {Katsnelson}}, \ and\ \bibinfo
  {author} {\bibfnamefont {S.}~\bibnamefont {Yuan}},\ }\bibfield  {title}
  {\enquote {\bibinfo {title} {Origin of the magic angle in twisted bilayer
  graphene from hybridization of valence and conduction bands},}\ }\href
  {\doibase 10.1103/PhysRevB.108.045138} {\bibfield  {journal} {\bibinfo
  {journal} {Phys. Rev. B}\ }\textbf {\bibinfo {volume} {108}},\ \bibinfo
  {pages} {045138} (\bibinfo {year} {2023})}\BibitemShut {NoStop}%
\bibitem [{\citenamefont {Grosso}\ and\ \citenamefont
  {Parravicini}()}]{GROSSO}%
  \BibitemOpen
  \bibfield  {author} {\bibinfo {author} {\bibfnamefont {G.}~\bibnamefont
  {Grosso}}\ and\ \bibinfo {author} {\bibfnamefont {G.~P.}\ \bibnamefont
  {Parravicini}},\ }\bibfield  {title} {\enquote {\bibinfo {title} {Chapter 12
  - {{Optical Properties}} of {{Semiconductors}} and {{Insulators}}},}\ }in\
  \href {\doibase 10.1016/B978-0-12-385030-0.00012-8} {\emph {\bibinfo
  {booktitle} {Solid {{State Physics}} ({{Second Edition}})}}},\ \bibinfo
  {editor} {edited by\ \bibinfo {editor} {\bibfnamefont {G.}~\bibnamefont
  {Grosso}}\ and\ \bibinfo {editor} {\bibfnamefont {G.~P.}\ \bibnamefont
  {Parravicini}}}\ (\bibinfo  {publisher} {{Academic Press}})\ pp.\ \bibinfo
  {pages} {529--576}\BibitemShut {NoStop}%
\bibitem [{\citenamefont {Giannozzi}\ \emph {et~al.}(2009)\citenamefont
  {Giannozzi}, \citenamefont {Baroni}, \citenamefont {Bonini}, \citenamefont
  {Calandra}, \citenamefont {Car}, \citenamefont {Cavazzoni}, \citenamefont
  {Ceresoli}, \citenamefont {Chiarotti}, \citenamefont {Cococcioni},
  \citenamefont {Dabo}, \citenamefont {Corso}, \citenamefont {de~Gironcoli},
  \citenamefont {Fabris}, \citenamefont {Fratesi}, \citenamefont {Gebauer},
  \citenamefont {Gerstmann}, \citenamefont {Gougoussis}, \citenamefont
  {Kokalj}, \citenamefont {Lazzeri}, \citenamefont {Martin-Samos},
  \citenamefont {Marzari}, \citenamefont {Mauri}, \citenamefont {Mazzarello},
  \citenamefont {Paolini}, \citenamefont {Pasquarello}, \citenamefont
  {Paulatto}, \citenamefont {Sbraccia}, \citenamefont {Scandolo}, \citenamefont
  {Sclauzero}, \citenamefont {Seitsonen}, \citenamefont {Smogunov},
  \citenamefont {Umari},\ and\ \citenamefont {Wentzcovitch}}]{Giannozzi2009}%
  \BibitemOpen
  \bibfield  {author} {\bibinfo {author} {\bibfnamefont {P.}~\bibnamefont
  {Giannozzi}}, \bibinfo {author} {\bibfnamefont {S.}~\bibnamefont {Baroni}},
  \bibinfo {author} {\bibfnamefont {N.}~\bibnamefont {Bonini}}, \bibinfo
  {author} {\bibfnamefont {M.}~\bibnamefont {Calandra}}, \bibinfo {author}
  {\bibfnamefont {R.}~\bibnamefont {Car}}, \bibinfo {author} {\bibfnamefont
  {C.}~\bibnamefont {Cavazzoni}}, \bibinfo {author} {\bibfnamefont
  {D.}~\bibnamefont {Ceresoli}}, \bibinfo {author} {\bibfnamefont {G.~L.}\
  \bibnamefont {Chiarotti}}, \bibinfo {author} {\bibfnamefont {M.}~\bibnamefont
  {Cococcioni}}, \bibinfo {author} {\bibfnamefont {I.}~\bibnamefont {Dabo}},
  \bibinfo {author} {\bibfnamefont {A.~Dal}\ \bibnamefont {Corso}}, \bibinfo
  {author} {\bibfnamefont {S.}~\bibnamefont {de~Gironcoli}}, \bibinfo {author}
  {\bibfnamefont {S.}~\bibnamefont {Fabris}}, \bibinfo {author} {\bibfnamefont
  {G.}~\bibnamefont {Fratesi}}, \bibinfo {author} {\bibfnamefont
  {R.}~\bibnamefont {Gebauer}}, \bibinfo {author} {\bibfnamefont
  {U.}~\bibnamefont {Gerstmann}}, \bibinfo {author} {\bibfnamefont
  {C.}~\bibnamefont {Gougoussis}}, \bibinfo {author} {\bibfnamefont
  {A.}~\bibnamefont {Kokalj}}, \bibinfo {author} {\bibfnamefont
  {M.}~\bibnamefont {Lazzeri}}, \bibinfo {author} {\bibfnamefont
  {L.}~\bibnamefont {Martin-Samos}}, \bibinfo {author} {\bibfnamefont
  {N.}~\bibnamefont {Marzari}}, \bibinfo {author} {\bibfnamefont
  {F.}~\bibnamefont {Mauri}}, \bibinfo {author} {\bibfnamefont
  {R.}~\bibnamefont {Mazzarello}}, \bibinfo {author} {\bibfnamefont
  {S.}~\bibnamefont {Paolini}}, \bibinfo {author} {\bibfnamefont
  {A.}~\bibnamefont {Pasquarello}}, \bibinfo {author} {\bibfnamefont
  {L.}~\bibnamefont {Paulatto}}, \bibinfo {author} {\bibfnamefont
  {C.}~\bibnamefont {Sbraccia}}, \bibinfo {author} {\bibfnamefont
  {S.}~\bibnamefont {Scandolo}}, \bibinfo {author} {\bibfnamefont
  {G.}~\bibnamefont {Sclauzero}}, \bibinfo {author} {\bibfnamefont {A.~P.}\
  \bibnamefont {Seitsonen}}, \bibinfo {author} {\bibfnamefont {A.}~\bibnamefont
  {Smogunov}}, \bibinfo {author} {\bibfnamefont {P.}~\bibnamefont {Umari}}, \
  and\ \bibinfo {author} {\bibfnamefont {R.~M.}\ \bibnamefont {Wentzcovitch}},\
  }\bibfield  {title} {\enquote {\bibinfo {title} {{QUANTUM ESPRESSO}: a
  modular and open-source software project for quantum simulations of
  materials},}\ }\href {\doibase 10.1088/0953-8984/21/39/395502} {\bibfield
  {journal} {\bibinfo  {journal} {J. Phys.: Condens. Matter}\ }\textbf
  {\bibinfo {volume} {21}},\ \bibinfo {pages} {395502} (\bibinfo {year}
  {2009})}\BibitemShut {NoStop}%
\bibitem [{\citenamefont {Giannozzi}\ \emph {et~al.}(2017)\citenamefont
  {Giannozzi}, \citenamefont {Andreussi}, \citenamefont {Brumme}, \citenamefont
  {Bunau}, \citenamefont {Nardelli}, \citenamefont {Calandra}, \citenamefont
  {Car}, \citenamefont {Cavazzoni}, \citenamefont {Ceresoli}, \citenamefont
  {Cococcioni}, \citenamefont {Colonna}, \citenamefont {Carnimeo},
  \citenamefont {Corso}, \citenamefont {de~Gironcoli}, \citenamefont {Delugas},
  \citenamefont {DiStasio}, \citenamefont {Ferretti}, \citenamefont {Floris},
  \citenamefont {Fratesi}, \citenamefont {Fugallo}, \citenamefont {Gebauer},
  \citenamefont {Gerstmann}, \citenamefont {Giustino}, \citenamefont {Gorni},
  \citenamefont {Jia}, \citenamefont {Kawamura}, \citenamefont {Ko},
  \citenamefont {Kokalj}, \citenamefont {K\"u\c{c}\"ukbenli}, \citenamefont
  {Lazzeri}, \citenamefont {Marsili}, \citenamefont {Marzari}, \citenamefont
  {Mauri}, \citenamefont {Nguyen}, \citenamefont {Nguyen}, \citenamefont {de-la
  Roza}, \citenamefont {Paulatto}, \citenamefont {Ponc\'e}, \citenamefont
  {Rocca}, \citenamefont {Sabatini}, \citenamefont {Santra}, \citenamefont
  {Schlipf}, \citenamefont {Seitsonen}, \citenamefont {Smogunov}, \citenamefont
  {Timrov}, \citenamefont {Thonhauser}, \citenamefont {Umari}, \citenamefont
  {Vast}, \citenamefont {Wu},\ and\ \citenamefont {Baroni}}]{Giannozzi2017}%
  \BibitemOpen
  \bibfield  {author} {\bibinfo {author} {\bibfnamefont {P.}~\bibnamefont
  {Giannozzi}}, \bibinfo {author} {\bibfnamefont {O.}~\bibnamefont
  {Andreussi}}, \bibinfo {author} {\bibfnamefont {T.}~\bibnamefont {Brumme}},
  \bibinfo {author} {\bibfnamefont {O.}~\bibnamefont {Bunau}}, \bibinfo
  {author} {\bibfnamefont {M.~Buongiorno}\ \bibnamefont {Nardelli}}, \bibinfo
  {author} {\bibfnamefont {M.}~\bibnamefont {Calandra}}, \bibinfo {author}
  {\bibfnamefont {R.}~\bibnamefont {Car}}, \bibinfo {author} {\bibfnamefont
  {C.}~\bibnamefont {Cavazzoni}}, \bibinfo {author} {\bibfnamefont
  {D.}~\bibnamefont {Ceresoli}}, \bibinfo {author} {\bibfnamefont
  {M.}~\bibnamefont {Cococcioni}}, \bibinfo {author} {\bibfnamefont
  {N.}~\bibnamefont {Colonna}}, \bibinfo {author} {\bibfnamefont
  {I.}~\bibnamefont {Carnimeo}}, \bibinfo {author} {\bibfnamefont {A.~Dal}\
  \bibnamefont {Corso}}, \bibinfo {author} {\bibfnamefont {S.}~\bibnamefont
  {de~Gironcoli}}, \bibinfo {author} {\bibfnamefont {P.}~\bibnamefont
  {Delugas}}, \bibinfo {author} {\bibfnamefont {R.~A.}\ \bibnamefont
  {DiStasio}}, \bibinfo {author} {\bibfnamefont {A.}~\bibnamefont {Ferretti}},
  \bibinfo {author} {\bibfnamefont {A.}~\bibnamefont {Floris}}, \bibinfo
  {author} {\bibfnamefont {G.}~\bibnamefont {Fratesi}}, \bibinfo {author}
  {\bibfnamefont {G.}~\bibnamefont {Fugallo}}, \bibinfo {author} {\bibfnamefont
  {R.}~\bibnamefont {Gebauer}}, \bibinfo {author} {\bibfnamefont
  {U.}~\bibnamefont {Gerstmann}}, \bibinfo {author} {\bibfnamefont
  {F.}~\bibnamefont {Giustino}}, \bibinfo {author} {\bibfnamefont
  {T.}~\bibnamefont {Gorni}}, \bibinfo {author} {\bibfnamefont
  {J.}~\bibnamefont {Jia}}, \bibinfo {author} {\bibfnamefont {M.}~\bibnamefont
  {Kawamura}}, \bibinfo {author} {\bibfnamefont {H.-Y.}\ \bibnamefont {Ko}},
  \bibinfo {author} {\bibfnamefont {A.}~\bibnamefont {Kokalj}}, \bibinfo
  {author} {\bibfnamefont {E.}~\bibnamefont {K\"u\c{c}\"ukbenli}}, \bibinfo
  {author} {\bibfnamefont {M.}~\bibnamefont {Lazzeri}}, \bibinfo {author}
  {\bibfnamefont {M.}~\bibnamefont {Marsili}}, \bibinfo {author} {\bibfnamefont
  {N.}~\bibnamefont {Marzari}}, \bibinfo {author} {\bibfnamefont
  {F.}~\bibnamefont {Mauri}}, \bibinfo {author} {\bibfnamefont {N.~L.}\
  \bibnamefont {Nguyen}}, \bibinfo {author} {\bibfnamefont {H.-V.}\
  \bibnamefont {Nguyen}}, \bibinfo {author} {\bibfnamefont {A.~Otero}\
  \bibnamefont {de-la Roza}}, \bibinfo {author} {\bibfnamefont
  {L.}~\bibnamefont {Paulatto}}, \bibinfo {author} {\bibfnamefont
  {S.}~\bibnamefont {Ponc\'e}}, \bibinfo {author} {\bibfnamefont
  {D.}~\bibnamefont {Rocca}}, \bibinfo {author} {\bibfnamefont
  {R.}~\bibnamefont {Sabatini}}, \bibinfo {author} {\bibfnamefont
  {B.}~\bibnamefont {Santra}}, \bibinfo {author} {\bibfnamefont
  {M.}~\bibnamefont {Schlipf}}, \bibinfo {author} {\bibfnamefont {A.~P.}\
  \bibnamefont {Seitsonen}}, \bibinfo {author} {\bibfnamefont {A.}~\bibnamefont
  {Smogunov}}, \bibinfo {author} {\bibfnamefont {I.}~\bibnamefont {Timrov}},
  \bibinfo {author} {\bibfnamefont {T.}~\bibnamefont {Thonhauser}}, \bibinfo
  {author} {\bibfnamefont {P.}~\bibnamefont {Umari}}, \bibinfo {author}
  {\bibfnamefont {N.}~\bibnamefont {Vast}}, \bibinfo {author} {\bibfnamefont
  {X.}~\bibnamefont {Wu}}, \ and\ \bibinfo {author} {\bibfnamefont
  {S.}~\bibnamefont {Baroni}},\ }\bibfield  {title} {\enquote {\bibinfo {title}
  {Advanced capabilities for materials modelling with {QUANTUM ESPRESSO}},}\
  }\href {\doibase 10.1088/1361-648X/aa8f79} {\bibfield  {journal} {\bibinfo
  {journal} {J. Phys.: Condens. Matter}\ }\textbf {\bibinfo {volume} {29}},\
  \bibinfo {pages} {465901} (\bibinfo {year} {2017})}\BibitemShut {NoStop}%
\bibitem [{\citenamefont {Perdew}\ \emph {et~al.}(1996)\citenamefont {Perdew},
  \citenamefont {Burke},\ and\ \citenamefont {Ernzerhof}}]{Perdew1996}%
  \BibitemOpen
  \bibfield  {author} {\bibinfo {author} {\bibfnamefont {J.~P.}\ \bibnamefont
  {Perdew}}, \bibinfo {author} {\bibfnamefont {K.}~\bibnamefont {Burke}}, \
  and\ \bibinfo {author} {\bibfnamefont {M.}~\bibnamefont {Ernzerhof}},\
  }\bibfield  {title} {\enquote {\bibinfo {title} {{Generalized gradient
  approximation made simple}},}\ }\href {\doibase 10.1103/PhysRevLett.77.3865}
  {\bibfield  {journal} {\bibinfo  {journal} {Phys. Rev. Lett.}\ }\textbf
  {\bibinfo {volume} {77}},\ \bibinfo {pages} {3865--3868} (\bibinfo {year}
  {1996})}\BibitemShut {NoStop}%
\bibitem [{\citenamefont {Pack}\ and\ \citenamefont
  {Monkhorst}(1977)}]{Pack1977}%
  \BibitemOpen
  \bibfield  {author} {\bibinfo {author} {\bibfnamefont {J.~D.}\ \bibnamefont
  {Pack}}\ and\ \bibinfo {author} {\bibfnamefont {H.~J.}\ \bibnamefont
  {Monkhorst}},\ }\bibfield  {title} {\enquote {\bibinfo {title} {{special
  points for Brillouin-zone integrations"-a reply}},}\ }\href {\doibase
  10.1103/PhysRevB.16.1748} {\bibfield  {journal} {\bibinfo  {journal} {Phys.
  Rev. B}\ }\textbf {\bibinfo {volume} {16}},\ \bibinfo {pages} {1748--1749}
  (\bibinfo {year} {1977})}\BibitemShut {NoStop}%
\bibitem [{\citenamefont {Marini}\ \emph {et~al.}(2009)\citenamefont {Marini},
  \citenamefont {Hogan}, \citenamefont {Gr\"uning},\ and\ \citenamefont
  {Varsano}}]{Marini2009}%
  \BibitemOpen
  \bibfield  {author} {\bibinfo {author} {\bibfnamefont {A.}~\bibnamefont
  {Marini}}, \bibinfo {author} {\bibfnamefont {C.}~\bibnamefont {Hogan}},
  \bibinfo {author} {\bibfnamefont {M.}~\bibnamefont {Gr\"uning}}, \ and\
  \bibinfo {author} {\bibfnamefont {D.}~\bibnamefont {Varsano}},\ }\bibfield
  {title} {\enquote {\bibinfo {title} {yambo: An ab initio tool for excited
  state calculations},}\ }\href {\doibase
  https://doi.org/10.1016/j.cpc.2009.02.003} {\bibfield  {journal} {\bibinfo
  {journal} {Comput. Phys. Commun.}\ }\textbf {\bibinfo {volume} {180}},\
  \bibinfo {pages} {1392--1403} (\bibinfo {year} {2009})}\BibitemShut {NoStop}%
\bibitem [{\citenamefont {Sangalli}\ \emph {et~al.}(2019)\citenamefont
  {Sangalli}, \citenamefont {Ferretti}, \citenamefont {Miranda}, \citenamefont
  {Attaccalite}, \citenamefont {Marri}, \citenamefont {Cannuccia},
  \citenamefont {Melo}, \citenamefont {Marsili}, \citenamefont {Paleari},
  \citenamefont {Marrazzo}, \citenamefont {Prandini}, \citenamefont
  {Bonf{\`{a}}}, \citenamefont {Atambo}, \citenamefont {Affinito},
  \citenamefont {Palummo}, \citenamefont {Molina-S{\'{a}}nchez}, \citenamefont
  {Hogan}, \citenamefont {Gr{\"{u}}ning}, \citenamefont {Varsano},\ and\
  \citenamefont {Marini}}]{Sangalli2019}%
  \BibitemOpen
  \bibfield  {author} {\bibinfo {author} {\bibfnamefont {D.}~\bibnamefont
  {Sangalli}}, \bibinfo {author} {\bibfnamefont {A.}~\bibnamefont {Ferretti}},
  \bibinfo {author} {\bibfnamefont {H.}~\bibnamefont {Miranda}}, \bibinfo
  {author} {\bibfnamefont {C.}~\bibnamefont {Attaccalite}}, \bibinfo {author}
  {\bibfnamefont {I.}~\bibnamefont {Marri}}, \bibinfo {author} {\bibfnamefont
  {E.}~\bibnamefont {Cannuccia}}, \bibinfo {author} {\bibfnamefont
  {P.}~\bibnamefont {Melo}}, \bibinfo {author} {\bibfnamefont {M.}~\bibnamefont
  {Marsili}}, \bibinfo {author} {\bibfnamefont {F.}~\bibnamefont {Paleari}},
  \bibinfo {author} {\bibfnamefont {A.}~\bibnamefont {Marrazzo}}, \bibinfo
  {author} {\bibfnamefont {G.}~\bibnamefont {Prandini}}, \bibinfo {author}
  {\bibfnamefont {P.}~\bibnamefont {Bonf{\`{a}}}}, \bibinfo {author}
  {\bibfnamefont {M.~O.}\ \bibnamefont {Atambo}}, \bibinfo {author}
  {\bibfnamefont {F.}~\bibnamefont {Affinito}}, \bibinfo {author}
  {\bibfnamefont {M.}~\bibnamefont {Palummo}}, \bibinfo {author} {\bibfnamefont
  {A.}~\bibnamefont {Molina-S{\'{a}}nchez}}, \bibinfo {author} {\bibfnamefont
  {C.}~\bibnamefont {Hogan}}, \bibinfo {author} {\bibfnamefont
  {M.}~\bibnamefont {Gr{\"{u}}ning}}, \bibinfo {author} {\bibfnamefont
  {D.}~\bibnamefont {Varsano}}, \ and\ \bibinfo {author} {\bibfnamefont
  {A.}~\bibnamefont {Marini}},\ }\bibfield  {title} {\enquote {\bibinfo {title}
  {{Many-body perturbation theory calculations using the yambo code}},}\ }\href
  {\doibase 10.1088/1361-648X/ab15d0} {\bibfield  {journal} {\bibinfo
  {journal} {J. Phys.: Condens. Matter}\ }\textbf {\bibinfo {volume} {31}},\
  \bibinfo {pages} {325902} (\bibinfo {year} {2019})},\ \Eprint
  {http://arxiv.org/abs/1902.03837} {arXiv:1902.03837} \BibitemShut {NoStop}%
\bibitem [{\citenamefont {Allen}\ \emph {et~al.}(2013)\citenamefont {Allen},
  \citenamefont {Berlijn}, \citenamefont {Casavant},\ and\ \citenamefont
  {Soler}}]{Allen2013}%
  \BibitemOpen
  \bibfield  {author} {\bibinfo {author} {\bibfnamefont {P.~B.}\ \bibnamefont
  {Allen}}, \bibinfo {author} {\bibfnamefont {T.}~\bibnamefont {Berlijn}},
  \bibinfo {author} {\bibfnamefont {D.~A.}\ \bibnamefont {Casavant}}, \ and\
  \bibinfo {author} {\bibfnamefont {J.~M.}\ \bibnamefont {Soler}},\ }\bibfield
  {title} {\enquote {\bibinfo {title} {Recovering hidden bloch character:
  Unfolding electrons, phonons, and slabs},}\ }\href {\doibase
  10.1103/PhysRevB.87.085322} {\bibfield  {journal} {\bibinfo  {journal} {Phys.
  Rev. B}\ }\textbf {\bibinfo {volume} {87}},\ \bibinfo {pages} {085322}
  (\bibinfo {year} {2013})}\BibitemShut {NoStop}%
\end{thebibliography}
%

\end{document}